\documentclass[preprint,longnamesfirst]{aastex}
\usepackage{multirow} 

\newcommand{\thisdir}{.}
\newcommand{\s}{\mathop{\rm s\,}\nolimits}

\newcommand{\yrs}{\mathop{\rm yrs\,}\nolimits}

\newcommand{\Ga}{\mathop{\rm Ga\,}\nolimits}

\newcommand{\km}{\mathop{\rm km\,}\nolimits}
\newcommand{\kpc}{\mathop{\rm kpc\,}\nolimits}
\newcommand{\Mpc}{\mathop{\rm Mpc\,}\nolimits}
\newcommand{\kps}{\mathop{\rm km/s\,}\nolimits}

\newcommand{\erg}{\mathop{\rm erg\,}\nolimits}
\newcommand{\keV}{\mathop{\rm keV\,}\nolimits}
\newcommand{\K}{\mathop{\rm K\,}\nolimits}

\newcommand{\Msun}{\mathop{\rm M_{\odot}\,}\nolimits}

\newcommand{\Lsun}{\mathop{\rm L_{\odot}\,}\nolimits}

\newcommand{\boltz}{\mathop{\rm k\,}\nolimits}
\newcommand{\G}{\mathop{\rm G\,}\nolimits}

\newcommand{\fig}{Fig.~\ref}

\newcommand{\tab}{Table~\ref}

\newcommand{\sect}{Sec.~\ref}
\newcommand{\sects}{Sections~\ref}

\newcommand{\eq}{Eq.~\ref}
\newcommand{\Hydra}{{\sc hydra}}

\newcommand{\app}{Appendix~\ref}

\newcommand{\expd}[1]{\times 10^{#1}}
\newcommand{\mean}[1]{\langle #1 \rangle}

\newcommand{\fraction}[2]{\mbox{\scriptsize$^{{#1}\!}/_{\!{#2}}$}}

\ifx\undefined\onequarter
 \newcommand{\onequarter}{\fraction{1}{4}}
\else
 \renewcommand{\onequarter}{\fraction{1}{4}}
\fi
\ifx\undefined\onethird
 \newcommand{\onethird}{\fraction{1}{3}}
\else
 \renewcommand{\onethird}{\fraction{1}{3}}
\fi
\ifx\undefined\onehalf
 \newcommand{\onehalf}{\fraction{1}{2}}
\else
 \renewcommand{\onehalf}{\fraction{1}{2}}
\fi
\ifx\undefined\twothirds
 \newcommand{\twothirds}{\fraction{2}{3}}
\else
 \renewcommand{\twothirds}{\fraction{2}{3}}
\fi
\ifx\undefined\threequarters
 \newcommand{\threequarters}{\fraction{3}{4}}
\else
 \renewcommand{\threequarters}{\fraction{3}{4}}
\fi

\ifx\undefined\fourthirds
 \newcommand{\fourthirds}{\fraction{4}{3}}
\else
 \renewcommand{\fourthirds}{\fraction{4}{3}}
\fi
\ifx\undefined\threehalfs
 \newcommand{\threehalfs}{\fraction{3}{2}}
\else
 \renewcommand{\threehalfs}{\fraction{3}{2}}
\fi
\ifx\undefined\threehalves
 \newcommand{\threehalves}{\threehalfs}
\else
 \renewcommand{\threehalves}{\threehalfs}
\fi

\newcommand{\figscale}{1}

\begin{document}

\title{Hierarchical clustering, the universal density profile, and the mass-temperature scaling law of galaxy clusters}

\author{Eric R. Tittley and H. M. P. Couchman\footnote{Present address: McMaster University, Department of Physics and Astronomy, Hamilton, Ontario, L8S 4M1, Canada}}
\email{etittley@astro.uwo.ca}
\email{couchman@physics.mcmaster.ca}
\affil{University of Western Ontario}
\affil{
Department of Physics and Astronomy, 
London, Ontario, N6A 3K7, 
Canada }

\begin{abstract}
The significance of hierarchical clustering on the density profile and mass-temperature scaling relation for galaxy clusters is examined using hydrodynamic N-body simulations. Clusters formed hierarchically are compared with clusters formed with the initial density fluctuations on sub-cluster scales removed via smoothing.

The universal profile, as described by \citeauthor{NFW95}, is not a by-product of hierarchical clustering.  It is found to fit the mean profiles of clusters formed both hierarchically and otherwise.  The Hernquist profile is also found to fit the data well.  The characteristic radius, $r_s$, moves outward from $0.1 R_{200}$ to $0.2 R_{200}$ when the initial substructure is eliminated.  Interior to $r_s$,  $\rho_{DM} \propto r^{-1.8}$, regardless of initial smoothing.  Exterior to this radius, the profile is marginally shallower in the non-hierarchical case, with $\rho_{DM} \propto r^{-2.4}$ compared with $\rho_{DM} \propto r^{-2.7}$.

The mass-temperature scaling relation maintains the form $T \propto M^{\twothirds}$, regardless of cluster formation method.  The normalisation varies at the 20\%\ level, which is at the level of the intrinsic scatter, with the non-hierarchical simulations producing the cooler clusters.
\end{abstract}


\shortcites{Adami98, Andreani99, BRFJ94, Carlberg97, Clowe98, Donahue98, Donahue99, EDFW85, Frenk98, Gioia99, Gioia90, Giraldi98, Henry97a, Komatsu99, MWNL99, Moore97, Moore98, Thomas98}

\section{INTRODUCTION}
\label{sec.Intro}
The largest virialised objects in the universe are galaxy clusters.  On scales larger than this, the quasi-linear collapse of large scale structure dominates.  Tthese clusters were initially identified optically from maps of galaxies which indicate galaxies tend to be found in together in groups, some of which contain many hundreds of galaxies.  Subsequently, these clusters of galaxies were found to contain a diffuse intergalactic medium composed of hot ($10^{8} \K \approx 10 \keV$) gas.  This gas is luminous ($\sim 10^{11} \Lsun$) in the x-ray regime and, consequently, observable from above the earth's atmosphere.  Two such satellites in particular, {\it Einstein} and {\it ROSAT}, have been sources for catalogues of x-ray luminosities \citep{Gioia90} as well as luminosity maps with sufficient resolution to map the coarse structure of these clusters \citep{MKU89,BRFJ94}.  This hot gas constitutes the bulk of the baryonic matter in galaxy clusters.

Satellites with instruments able to take spectral data in the x-ray band, most notably the {\it ASCA} satellite, have coarsely resolved the spatial temperature distribution of this gas (see \citet{MV}, for references).  A new breed of x-ray telescopes including {\em Chandra}, {\it XMM}, and {\it Astro-e} will be able to better map the temperature and mass distributions of clusters \citep{Schind99}, potentially to a level which can discriminate competing cosmological models.

Being so luminous, galaxy clusters are observable to great distances and, consequently, at earlier stages of their evolution. This permits tests of many cosmological parameters \citep{Henry97a,Donahue98,Donahue99,Gioia99}.  Galaxy clusters are observable in ways other than their x-ray signature and population of galaxies.  In the microwave spectrum, there is the temperature decrement imposed on the cosmic microwave background via the Sunyaev-Zeldovich effect by which electrons in the hot gas alter the energy spectrum of the CMB photons \citep[see][for examples]{CJG,Komatsu99,Andreani99}.  Optical imaging of the clusters also provides independent measurements of the total mass of the clusters via the gravitational lensing of background galaxies \citep[see][for examples]{Clowe98,MA99}, though the method still has details to be clarified \citep{MWNL99}.

In the standard model of cosmological structure formation, clusters are the most recently relaxed large-scale objects.  Consequently, though they are relaxed, they potentially carry with them information related to their formation.  Larger-scale structure is in a state of linear or quasi-linear evolution and, hence, still responding to density perturbations that have existed since early times.  Smaller virialised objects such as galaxies and stars individually contain little or no manifestation of the character of the initial density perturbations from which they formed.  Having had ample time to dynamically relax, their present evolution is dominated by the physics of stellar evolution. Clusters are representative of structure on the intermediate scale.

There is ample evidence that the dominant form of matter in the universe is unseen and not baryonic in nature.  This dark matter is fundamental to the standard model.  Though the presence of dark matter may be inferred from observational data, by its nature it cannot be presently readily observed directly.  Numerical simulations which model this collisionless dark matter provide a powerful tool to achieve the goal of understanding the history and dynamics of structure formation. Two recent results stemming from numerical simulations have added to our understanding of the morphology of galaxy clusters as well as provided a tool for inferring the total mass of clusters from their measured temperatures.  These are the universal profile of dark matter and the mass-temperature scaling relation.

\subsection{A universal density profile}
\label{sec.Intro.UnivProf}
The works of \citet{NFW95} and \citet{NFW96a} have built a case for a common density profile for the dark matter in galaxy clusters.  This density profile is found to fit clusters spanning a large range of masses.  It is contended that this `universal' profile has only one free parameter corresponding to the density at cluster formation \citep{NFW96b}.  However, the form of the profile is derived from numerical simulations of hierarchically formed clusters.  There is evidence that it is the hierarchical nature of the cluster formation itself that is responsible for the universal profile \citep{SW}.  Determining the validity of this would be valuable to understanding the true breadth of its universality.

\subsection{The mass-temperature scaling relation}
\label{sec.Intro.MT}
Estimations of the total mass of the clusters have been made using the observed x-ray luminosity (and hence, baryon) distribution and the assumption of hydrostatic equilibrium using so-called $\beta$-model fits \citep{FG83,JF84}.  \citet{EMN} describes a correlation between the total mass of clusters of galaxies and the mass-weighted mean temperature of the hot, x-ray emitting gas in the interior of the cluster.  This would provide an independent method of measuring the total masses of the clusters, requiring only a measurement of the temperature of the gas.  This relationship is a natural result of the hydrostatic state of the gas.  However, derivation of this relationship uncovers a dependency on both the gas density and temperature profiles.  Hence, a sensitivity of these profiles to the cosmogony would weaken the relation's utility.  Understanding the behaviour of the mass-temperature scaling relation on the cosmogony is essential to the confidence level put in the application of the relation to real data.

\subsection{Hierarchical clustering}
\label{sec.Intro.HierClust}
In hierarchical clustering, the largest structures forming at a given time do so via the amalgamation of many smaller structures which have formed at an earlier time.  This is owing to the form of the initial density perturbation spectrum in which small scale perturbations have higher initial amplitudes than large scale. In contrast, non-hierarchical clustering involves structure formation from the collapse of large structures with smooth density distributions.  Though the details are still not clear, the results of numerical simulations compared with observations support the theory that we live in a universe in which structure is formed hierarchically.  The degree to which the hierarchical nature affects galaxy clusters is not entirely clear.

  
In this paper, the issue of whether the mass-temperature scaling relationship and the common density profile are universal between hierarchical and non-hierarchical cluster formation scenarios is explored.  By using these extreme cases, the significance of hierarchical clustering itself will be determined.  Through inference, their dependence on the presumed cosmological model will also be illuminated; if they are found to be independent on hierarchical clustering, then it would be expected that the precise form of hierarchical clustering would also be irrelevant. 

The analysis will concentrate not on the properties of individual clusters, but on the global mean properties of scaled quantities.

The layout of the paper is as follows. In \sect{Sec.Sims}, the simulations are described. A brief description of the analysis methods is given in \sect{Sec.Analysis}. Verification that the clusters extracted are in hydrostatic equilibrium is made in \sect{sec.hydrostatic}. The dark matter density profile is examined in \sect{sec.UniversalProfile}.  The derivation of the mass-temperature relationship and the results relevant to clusters are presented in \sect{sec.MT}. The conclusions are summarised in \sect{sec.Conclusions}.

\section{SIMULATIONS}
\label{Sec.Sims}
\renewcommand{\thisdir}{Simulations}
\subsection{Numerics}
\label{sec.Sims.Numerics}
The N-body ${\rm AP}^3{\rm M}-{\rm SPH}$ code, \Hydra\ \citep{CTP} was used for all simulations.  This is a multi-level Adaptive Particle-Particle Particle-Mesh (${\rm AP}^3{\rm M}$) N-body code with gas dynamics simulated by the Lagrangian Smooth Particle Hydrodynamics (SPH) method.  The gravity is calculated using a particle-mesh scheme for the large scale gravitational fields.  Short-range forces are calculated by summing particle-particle forces.  Isolated regions of high number density have sub-grids adaptively placed around them, allowing the efficient particle-mesh method to be used locally.  As such, it is well suited for the large dynamic range of scales involved in cluster studies.  A comparison of the leading hydrodynamic codes, including \Hydra, designed for studying cosmological scenarios, finds general consistency among the codes \citep{Frenk99}.

\subsection{Cosmology}
\label{sec.Sims.Cosmology}
All simulations assumed the flat cosmology given in \tab{tab.Sim.cosmology}.
The matter was evolved in a box with $40 h^{-1} \Mpc$ sides in co-moving coordinates, with the Hubble expansion constant, $H_o=h100\kps$.  This permits the formation of a sufficient number of clusters for statistical purposes.  A box of larger dimensions would impose too great a penalty on the spatial and mass resolution of the simulation. A smaller box would lack the long-wavelength perturbations required to produce large clusters. The simulation had periodic boundary conditions.

Cooling is neglected.  The cooling time for the bulk of the cluster gas is estimated to be well over the age of the universe.  The peak luminosity is on the order of $10^{44} \erg \s^{-1}$.  The thermal energy found in the gas of a system containing $10^{14} \Msun$ of gas at $5 \keV$ is on the order of $10^{63} \erg$ giving a cooling time, $t_{cool}=10^{19} \s \approx  3\expd{11}\yrs$.  There is insufficient resolution due to the limit set by the gravitational softening parameter to properly model the cooling flows which are inferred in the inner $200 \kpc$ of clusters \citep{AF97}.  It would require a great deal more resolution in order to model these flows correctly owing to the overcooling induced by the inability of SPH to properly calculate the gas densities in the cores of clusters.  As well, when modelled with sufficient resolution, the cooling flows simulated are much greater than those observed, likely due to a lack of feedback mechanisms such as supernova energy input which reheats the gas \citep{SO98}.  In any case, the amount of gas inferred to have cooled in these flows is on the order of $10^{11} \Msun$, which is not significant to the larger scale distribution of gas in the cluster.  

\subsection{Initial conditions}
\label{sec.Sims.InitialConditions}
The initial density perturbations were established by displacing the particle positions from a uniform cubic grid using, in the standard way \citep{Efstathiou85}, the Zel'dovich approximation for growth of density perturbations in the linear regime \citep{Zeld70}.  This method first creates a representation of the density perturbations in Fourier k-space using a supplied power spectrum form supplemented with Gaussian random fluctuations.  This is then transformed into an initial density field.  The gravitational forces felt by the set of particles, distributed at the nodes of the uniform mesh, by this density field are calculated.  The particles are then displaced in the direction of their respective forces an amount proportional to the force.  This produces a distribution of particles with a density field following the previously calculated density field.

The initial redshift for the simulations is $z_{initial}=75$. This was chosen to keep the maximum displacement incurred during the establishment of the density field to less than \onehalf\ the initial grid spacing.  This makes negligible the errors incurred in using the linear Zel'dovich approximation to a non-linear system.  The initial power spectrum of the density fluctuations follows a power-law of $n=-1$, $P(k) \propto k^{-1}$.

In order to ascertain the effects hierarchical clustering has on the matter, runs with initial high-spatial-frequency density perturbations were compared to runs in which these density perturbations were suppressed. The perturbation suppression was accomplished via two methods.

For the first, the initial density field was convolved with a tophat.  This modification to the power spectrum takes the form $P^\prime(k) = P(k)W(kr_{smooth})$ where $W(x)$ is the tophat window function given by,
\begin{equation}
W(x)=\frac{3(\sin x - x \cos x)}{x^3}.
\label{eq.Sim.window}
\end{equation}
A choice of $r_{smooth} = 7 h^{-1} \Mpc$ for the smoothing length was selected to suppress those perturbations of spatial size less than the size of a cluster formation region.  Consequently, the clusters in this scenario form from the uniform collapse of structures of a size on the scale of or larger than those that formed the clusters seen at the present.  To span the regimes, a set of initial conditions smoothed with $r_{smooth} = 3 h^{-1} \Mpc$ was also evolved. In this case, the early generations of structure formation would be suppressed, but the largest objects at the final epoch would have formed from the merger of a few objects.

\begin{figure}
\epsscale{\figscale}
\plotone{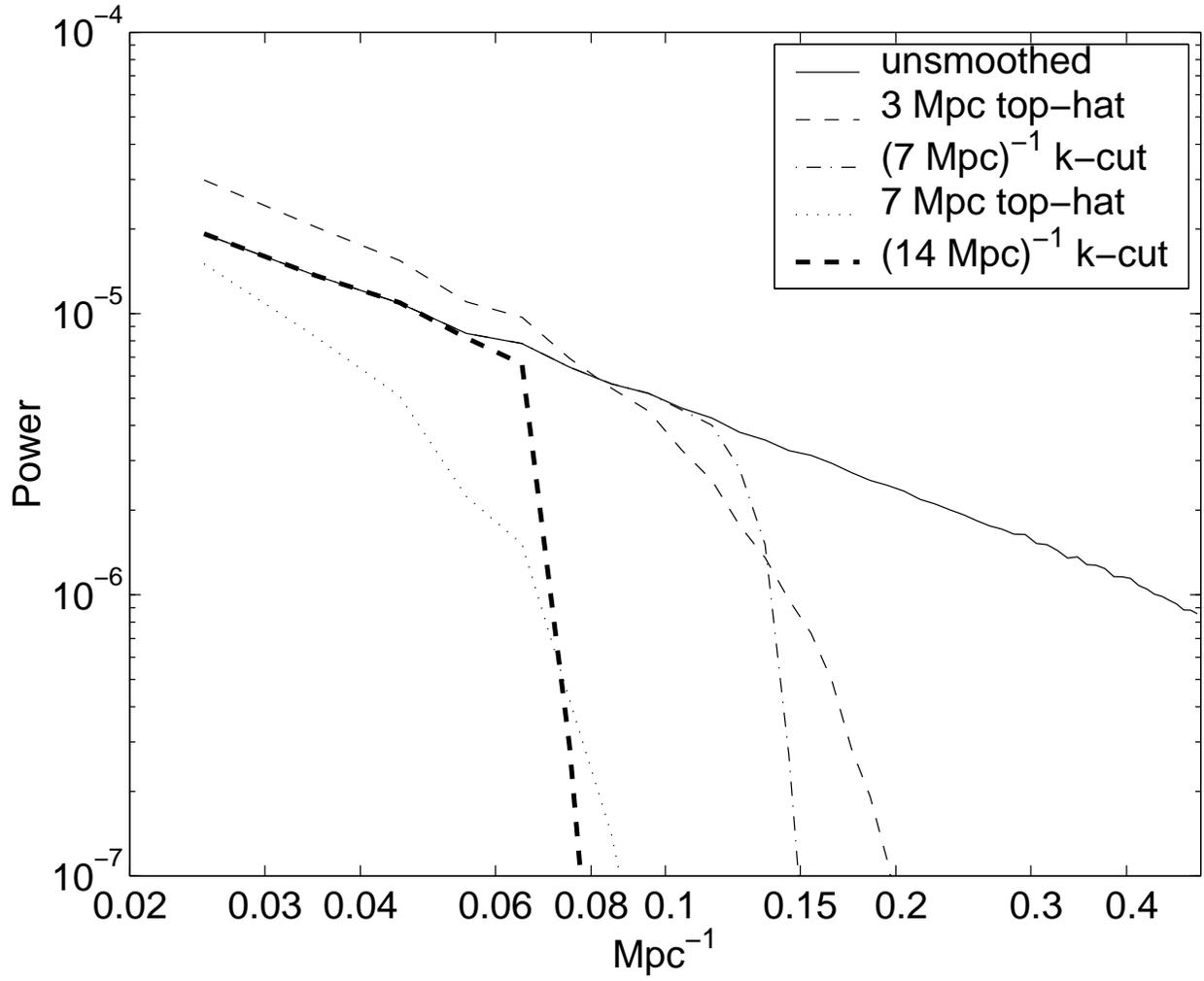}
\caption{The initial power spectra of the simulations.}
\label{fig.Sims.InitialPS}
\end{figure}

The second method for smoothing the initial power spectrum is a low-bandpass filter applied to the power spectrum of the density-perturbations before the particles are displaced.  This filter is of the form
\begin{equation}
P^\prime(k) = P(k)e^{-\left(\frac{kr_{smooth}}{2\pi}\right)^n},
\label{eq.Sim.freq-cut}
\end{equation}
where $n$ is some positive integer which controls the sharpness of the cutoff which was set to $n=16$.  Two sets of initial conditions for which $r_{smooth}=7 h^{-1} \Mpc$ and $14 h^{-1} \Mpc$ were created using this method.  The power spectra for the initial conditions created with the top-hat smoothing as well as the frequency cutoff are illustrated in \fig{fig.Sims.InitialPS}.

\begin{figure}

\centering
\includegraphics[width=0.4\textwidth]{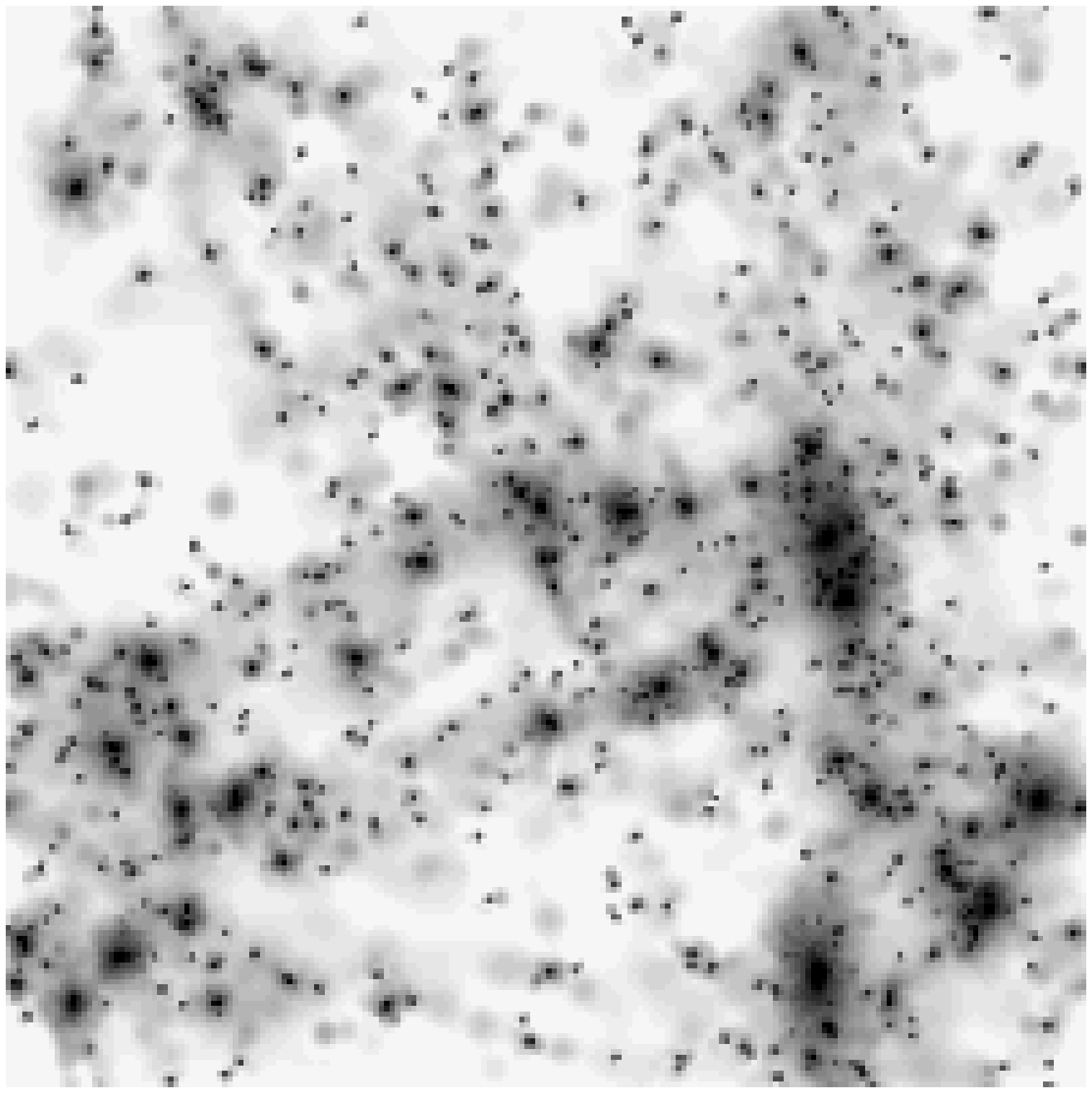}
\begin{minipage}[b]{0.4\textwidth}
\caption{
The projected gas densities of the simulation boxes at $t=1$ ({\it i.e.}, present).  Top left plate is evolved from the unsmoothed initial condition (IC); middle left from the $3 \Mpc$ top-hat smoothed IC; middle right from the $(7 \Mpc)^{-1}$ cut-off filtered IC; bottom left from the $7 \Mpc$ top-hat smoothed IC; and bottom right from $(14 \Mpc)^{-1}$ cut-off filtered IC.  Darker shades indicate higher densities, in log scale.}
\label{fig.Sims.density}
\end{minipage}
\includegraphics[width=0.4\textwidth]{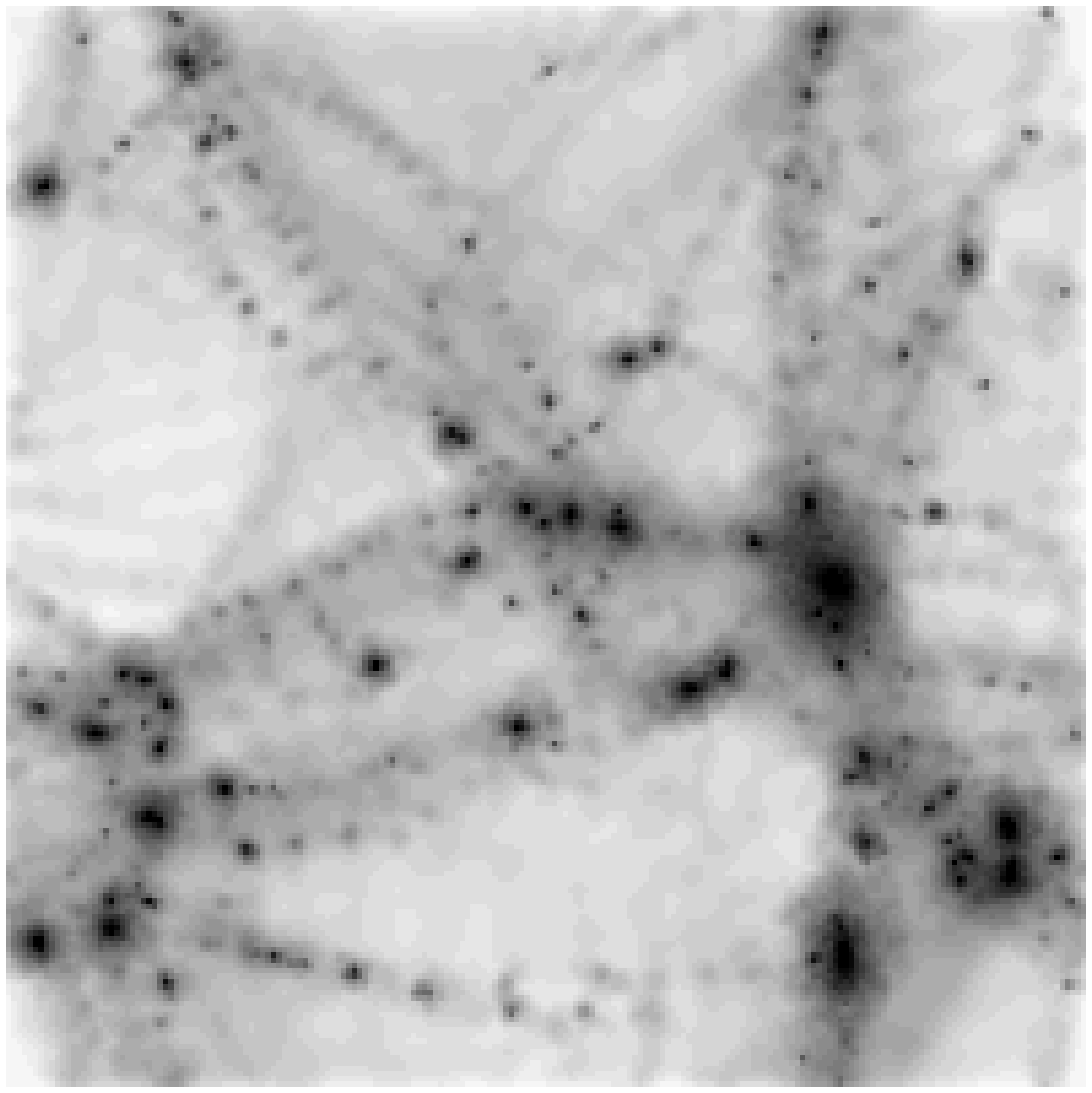}
\includegraphics[width=0.4\textwidth]{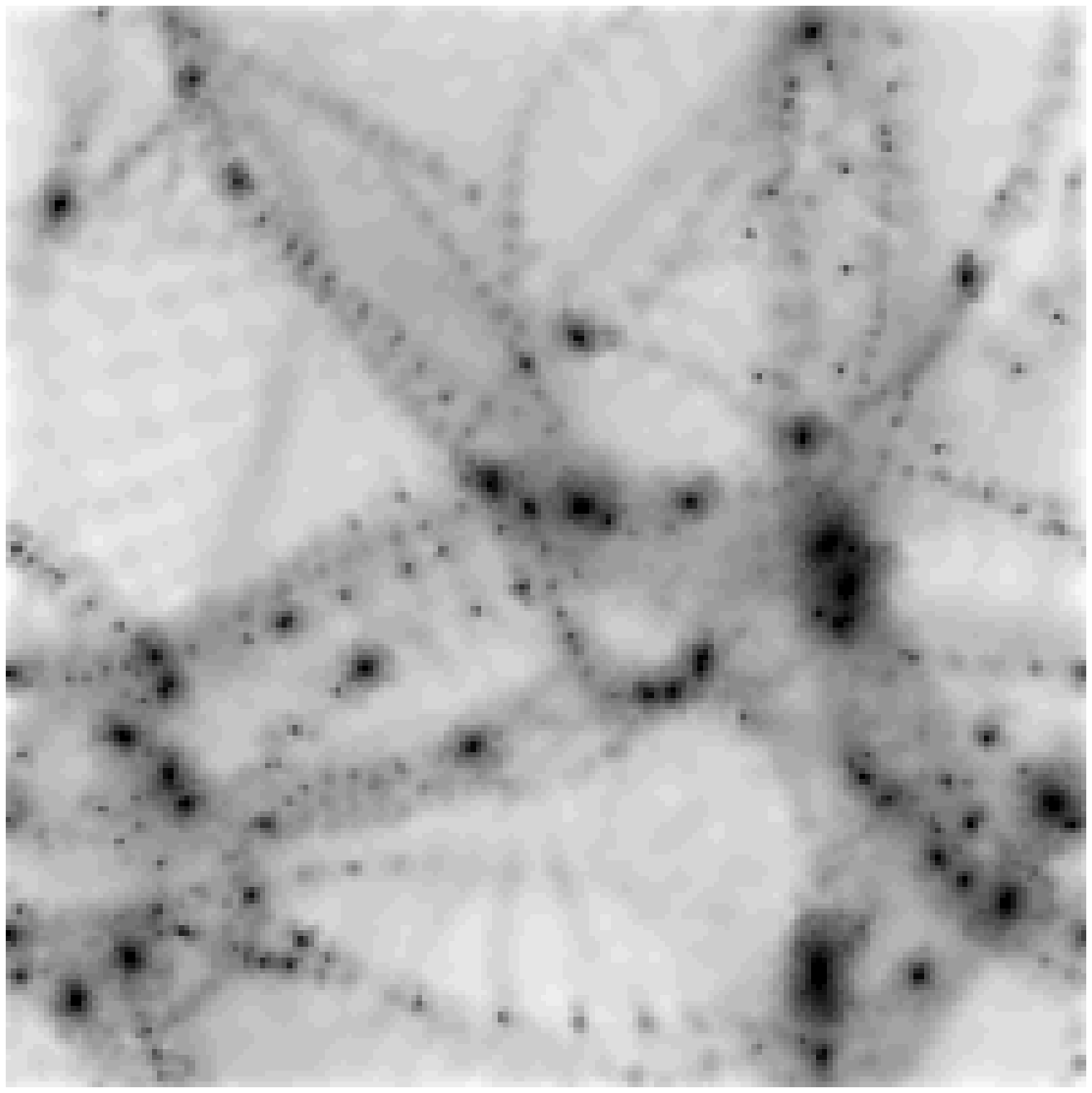} \\
\includegraphics[width=0.4\textwidth]{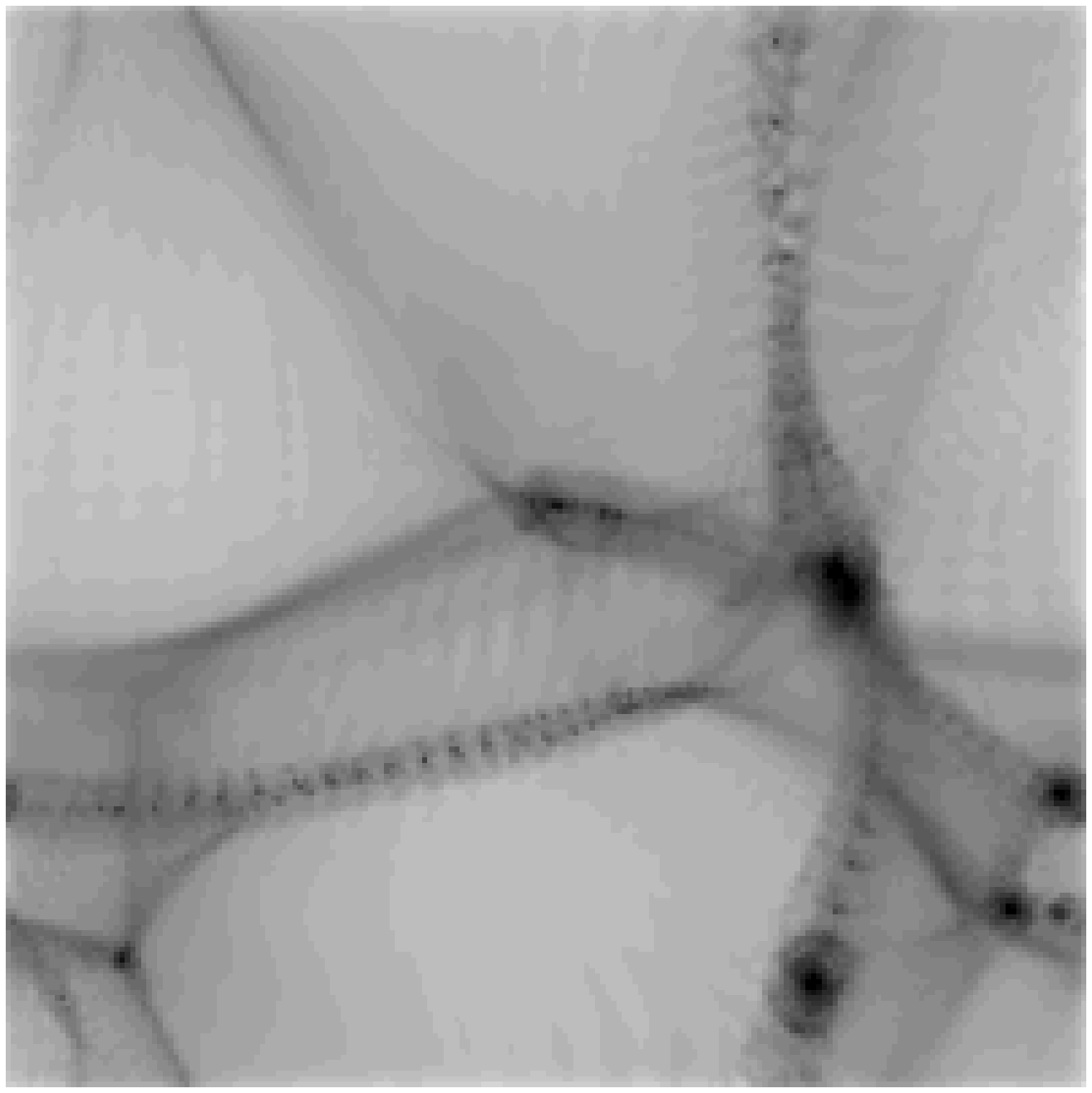}
\includegraphics[width=0.4\textwidth]{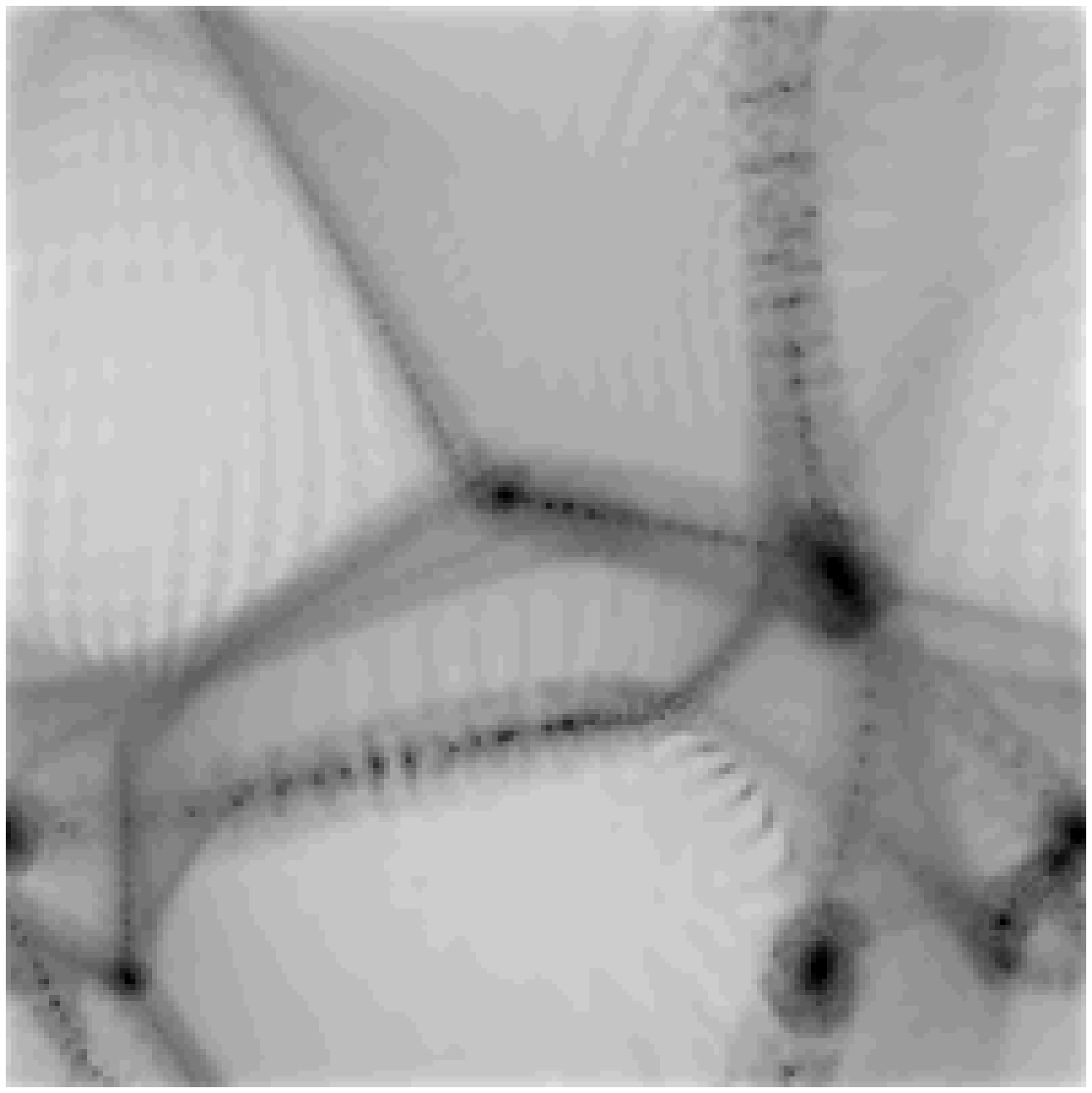}
\end{figure}

The first minimum in $W(x)$ is at $k\approx \frac{5}{7}\frac{2\pi}{r_{smooth}}$. Hence, tophat smoothing over a radius of $7 h^{-1} \Mpc$ removes more power at frequencies $k < 2\pi/r_{smooth}$ than a frequency-cutoff filter using the same value of $r_{smooth}$. The significance of this is made apparent by noting that the frequency-cutoff filter is essentially a convolution in real space of the density distribution with the window function, $W(x)$.  The $W(x)$ function can crudely be thought of as a top-hat with a radius given by the position of the first minimum.  This implies a frequency cutoff of $2\pi(7 h^{-1} \Mpc)^{-1}$ is roughly comparable to smoothing with a top-hat of radius $5 h^{-1} \Mpc$. Our choice of $r_{smooth}$ for the frequency filtering reflects this, with a frequency cutoff of $2\pi(7 h^{-1} \Mpc)^{-1}$ producing an initial density distribution roughly matching that produced by $3 h^{-1} \Mpc$ top-hat smoothing.  Similarly, the cutoff of $2\pi(14 h^{-1} \Mpc)^{-1}$ produces results similar to the $7 h^{-1} \Mpc$ top-hat.  This is verified at a visual level in the distribution of matter at the end of the simulations (\fig{fig.Sims.density}). For brevity, subsequent references to the scale of frequency cutoffs will omit the $2\pi$ term, which may be assumed.

Since the top-hat smoothing removes power at all frequencies except 0, the power on all scales was increased during the $3 h^{-1} \Mpc$ top-hat smoothing to maintain the same RMS fluctuations on scales of $8h^{-1}\Mpc$ ($\sigma_8$) as found in the unsmoothed set of initial conditions.  This was not done for the $(14 h^{-1} \Mpc)^{-1}$ frequency filtering, as this removes power up to scales larger than $8 h^{-1} \Mpc$, nor was it done for the $7 h^{-1} \Mpc$ top-hat filter for which first minimum in k-space is too close to the scale of $(8 h^{-1} \Mpc)^{-1}$.  Frequency filtering over $(7 h^{-1} \Mpc)^{-1}$ does not require this correction.

To examine the effects of resolution, the runs were performed at two resolutions. In one set, $64^3$ particles each of gas and of a collisionless dark matter component were evolved.  In the other set, $32^3$ particles of each were used. The gravitational softening was $0.02 h^{-1} \Mpc$ and $0.04 h^{-1} \Mpc$ for the respective runs. The details of the simulations are given in \tab{tab.Sim.simulations}. The difference in structure at $t=1$ (i.e., the present) is clearly illustrated in \fig{fig.Sims.density} which maps the projected gas density in the each of the simulated volumes.

\subsection{Two-body heating}
\label{sec.Sims.2bodyHeating}
The effect of two-body interactions is minimised by the gravitational softening parameter, $\epsilon$.  Without a gravitational softening parameter, gravitational interactions heat the gas via an exchange of energy between the dark matter and gas phases during short-range encounters of pairs of these particles \citep{SW97}.  If the mean interparticle spacing is on the order of or greater than $\epsilon$, then the time, $t_{2-body}$, required to heat the gas to its temperature, $T$, via this process is given by,
\begin{equation}
\left( \frac{t_{2-body}}{10^{10} \yrs} \right) =
160 \frac{ \left(\frac{T}{10^6\K}\right)^{\threehalfs} }
{ \left(\frac{\ln\Lambda}{5}\right)
  \left(\frac{m_{DM}}{10^{10}\Msun}\right)
  \left(\frac{m_{DM}}{m_{gas}}\right)^{\onethird}
  \left(\frac{\rho_{DM}}{\rho_c}\right)
  h^2}
\label{eq.Sims.t_2body}
\end{equation}
where $\ln\Lambda$ is the Coulomb logarithm (3-7 for most simulations), $m_{DM}$ and $m_{gas}$ are the respective masses per particle for the dark matter and gas, $\rho_{DM}$ is the dark matter mass density, and $\rho_c$ is critical density.

Without a softening term in the gravitational forces, two-body interactions would heat the gas in the centre of simulated clusters to $10^7 \K$ in $0.5 \Ga$.  Indeed, it can be shown for the simulations presented here, in the absence of a gravitational softening term, two-body heating would be significant for radii less than $\sim 0.17 \Mpc$ for the simulations using $2\times 64^3$ particles and $\sim 0.5 \Mpc$ for those using $2\times 32^3$. In the dense cores, where this process is most significant owing to the dependency, $t_{2-body} \propto \rho_{DM}^{-1}$, there are on the order of 70 particle within a softening length, $\epsilon$, of any other particle. This obliviates the problem, there. As the density drops, however, this number drops.  Again, using the density profile information described in \sect{Sec.Profiles.ResultsDM}, it may be found that the mean number of particles within $\epsilon$ falls to unity at a radius of $\sim 0.1 \Mpc$, independent of the spatial resolution of the simulations, since $\epsilon$ scales with spatial resolution in these simulations. This implies that two-body heating is probably a factor in the regimes of $0.1$ to $0.17 \Mpc$ for the high-resolution simulations and  $0.1$ to $0.5 \Mpc$ for the lower.  In the outer halo, $\rho_{DM}$ falls off faster with radius than does the factor, $T^{\threehalfs}$, so two-body heating does not become a factor except, perhaps, for the cold void particles.

\section{ANALYSIS}
\label{Sec.Analysis}
\subsection{Cluster selection}
\label{sec.Analysis.ClusSelection}
Clusters were selected by a tomographic deprojection method.   The gas distribution, extended over its smoothing radius, was projected on to three planes: the x-y, the x-z, and the y-z.  The resolution of these projections is $\approx 100\kpc$.  Peaks were found in these three planes which were then tomographically deprojected to recover the (x,y,z) coordinates of the density maximums.  The deprojection method involves the following steps:
\begin{enumerate}
\item Match the peak positions in the x domain (within some tolerance) for the x-y and x-z projections to find a list of possible y-z positions.
\item Match the y-z positions postulated from the previous step with the y-z positions found from the peaks in the y-z projection.
\item For each of the matches, use the x position from step 1) to get an x-y-z position.
\end{enumerate}
This method is comparable to projecting the gas densities onto a 3-dimensional mesh of resolution (box size)$/L$, but requires $3L^2$ elements of information instead of $L^3$ which, beyond the memory requirement savings, speeds up the peak-finding algorithm.

Cluster searching using the friends-of-friends method ({\sc fof v1.1}~\footnote{see  http://www-hpcc.astro.washington.edu/tools/FOF/}) did not give satisfactory results.  This method links together particles closer than a distance given by a linking length and then associates to a particle all its neighbour particles as well as their neighbours and so on.  The association is then called a cluster.  Close clusters are often connected by bridges allowing the association to `percolate' through.  As such, the grouping of cluster members was found to be too sensitive to the linking length parameter.  For a choice of a linking length suitable for the largest clusters, the smallest clusters were not properly recovered.  Another popular routine, {\sc skid}~\footnote{see  http://www-hpcc.astro.washington.edu/tools/SKID/} (based on {\sc denmax}), was tested.  This routine `freezes' the particles, then allows them to gravitate together to form tighter groups, essentially severing the bridges between clusters.  A friends-of-friends procedure is then allowed to create the list of associations.  Though the routine {\sc denmax} performed better than {\sc fof}, particularly for the $2\times 32^3$ simulation, the amount of CPU time required for the high-resolution simulation was unacceptable, taking days instead of hours.

The centres of the density peaks are further refined by cutting out a small sphere of particles, centred on the estimated density peak, and finding the mean position of the particles with the highest densities calculated previously by the SPH algorithm.

Overdensity radii, $R_{\bar\delta}$, were calculated for the clusters for overdensities of $\bar\delta \equiv \frac{\rho}{\rho_c} = 200$ and 500.  An overdensity radius is simply the radius from the centre of a cluster within which the ratio of the mean density to the critical density of the universe is equal to some value, referred to as the overdensity.  Though spherical symmetry is not required, the derivation holds more relevance for spherically symmetric systems. The expected radius of virialisation derived from analytic models involving simple spherically symmetric collapse corresponds to that of an overdensity of 178 (i.e., $R_{178}$).  However, the overdensity radius, $R_{\bar\delta}$, is not particularly sensitive to $\bar\delta$.  The overdensity within the calculated overdensity radius is accurate to a factor of $N^{-1}$ where $N$ is the number of particles within $R_{\bar\delta}$.  With the overdensity radii, there is a corresponding overdensity mass, $M_{\bar\delta}$, which is the mass contained within $R_{\bar\delta}$.

A lower limit to the size of the clusters was set by the requirement that each cluster, within the overdensity radius of $R_{500}$, have at least 300 gas particles and 300 dark matter particles.  This ensures the densities are calculated correctly (see \app{sec.Tech.Grad}).

The details of the sets of clusters found using this method are given in \tab{tab.analysis.clusters}.

A similar procedure was done using the dark matter mass distribution.  The tomographic deprojection step for finding the clusters first required SPH-like density estimates be made for the dark matter and appropriate smoothing lengths be calculated in order to project the masses onto the x-y, x-z, and y-z planes. Little difference was found in the positions or number of the final clusters selected when the dark matter distribution was used instead of that of the gas.

\subsection{Cluster profiles}
\label{sec.Analysis.Profiles}
Profiles of various parameters were calculated for the clusters by summing the contributions of the particles falling in radial bins centred on the clusters.  The bins were separated exponentially.  Found for each of the dark matter and gas components were the number of particles, the volume-weighted mean density, and the mass-weighted mean thermal energy.  The innermost bin was selected to be at a radius greater than that which encloses 300 particles to ensure that the gas dynamics well-models the density gradient (see \app{sec.Tech.Grad}).

The thermal noise of the SPH particles provides a negligible contribution ($1\%$) towards the thermal energy of gas particles in the haloes.  This contribution is ignored in all cases.

Mean profiles were calculated using the radially binned profiles described previously.  The individual cluster profiles were scaled radially by the overdensity radius, $R_{200}$, as well as weighted by their respective cluster mass.  For the density profiles, this weighting was found to have only a small effect, however, validating the use of mean profiles.

Density and temperature profile forms were fit to either the mean profiles or to the particle distributions of density or temperature versus radius.  Details of the fits are given in \sects{Sec.Profiles.ResultsDM},~\ref{sec.MT.GasDensityProfile} and~\ref{sec.MT.TauProfile}.  In all cases, the fits were found by minimising the $\chi^2$ with the variance derived from the distribution of values in radial bins.  Again, the distances were scaled by the overdensity radius, $R_{200}$.  When fitting to the particle distributions of density or temperature versus radius, all the data from selected clusters in a given simulation were used simultaneously.

Errors for the fits to the density or temperature versus radius distributions were determined using a bootstrap method.  In this method, one half of the data are fit at a time, giving a set of parameters to the fit.  This is done a multitude of instances for the same data, selecting a separate random sample of points in the distribution for each instance.  These fits produce a set of values for each parameter to the fit, from which a mean and deviation is derived.

It is the SPH-calculated density that is used when the particle distribution of density versus radius is fit.  Both the dark matter and gas particles had their densities calculated more precisely by performing an SPH summation over a value for the smoothing parameter,  $h$, such that $2h$ encompasses exactly $N_{SPH} = 32$ particles.  For the gas, this primarily affects the low and high density regimes since during the simulation there is a lower and upper bound set on $h$ by the gravitational softening length and the spacing of the mesh used to facilitate neighbour searching, respectively.  The densities were calculated without a self-density term to provide a volume-weighted average local density at each of the particle positions.

\section{HYDROSTATIC EQUILIBRIUM}
\label{sec.hydrostatic}
The derivation of the mass-temperature scaling relation, as well as the $\beta$-model method for estimating the mass of galaxy clusters require that the clusters be in hydrostatic equilibrium within some distance from the cluster centre.  This can be tested in the numerical simulations.  Integration of the equation of hydrostatic equilibrium produces an estimate for the gas pressure which can be compared with the actual gas pressure profile.  This integral is given by,
\begin{equation}
P_g(r) = \int_r^{\infty} \mean{\rho_g(r')} \frac{\G M(r')}{r'^2} dr'
\label{eq.hydro_eq.P_g}
\end{equation}
where $M(r)$ is the total mass interior to the shell of radius $r$ and $\mean{\rho_g(r)}$ is the mean gas density at $r$. The calculated value, $P_g(r)$, is the gas density at $r$ required to satisfy the condition of hydrostatic equilibrium. In practice, integration to $r \rightarrow \infty$ is both impractical, due to the finite box size, and inappropriate, since the cluster is not isolated. Instead, a radius at which the presence of nearby clusters influences the profiles is chosen.  This is always at a distance such that the contribution to the integral will be minimal.

The analysis described above was performed on the sample of cosmological clusters described in \sect{Sec.Sims}.  The results for the hierarchically formed clusters, shown in \fig{fig.hydrostatic.hydro_eq.run101}, indicate that the clusters have mean radial pressure profiles that are consistent with the condition of hydrostatic equilibrium. This is particularly true within the virial radius indicated by $R_{200}$.  The same is true for clusters formed non-hierarchically (\fig{fig.hydrostatic.hydro_eq.run103}), for which we used the data for the clusters formed from initial conditions smoothed by a $7h^{-1} \Mpc$ top-hat as the sample.  Consequently, it may be accepted that the clusters in this study are approximately hydrostatic out to their virial radii.

\renewcommand{\thisdir}{Profiles}
\section{UNIVERSAL PROFILE}
\label{sec.UniversalProfile}
Numerical simulations have indicated that, for a Cold Dark Matter (CDM) model, the dark matter density profile of galaxy clusters may be fit by a form with only one free parameter set by the mean density of the universe at the time of collapse \citep{NFW96a}.  The existence of such a universal density profile (or NFW profile) would be a useful tool as well as a test for theories of cluster formation.  The applicability of the universal profile to clusters formed in other cosmological models would strengthen its use as a tool for observers as well as clarify its ability to discriminate among models.  Expectations for the shape of the density profile have been made which can explain the existence of a universal profile and support the NFW form \citep{EC97,Pad96b}.  Some of these depend on hierarchical clustering explicitly.

The dependency of a universal density profile on hierarchical clustering is explored in this section.  First, some of the possible forms of the density profile will be examined.  After this, the results of fitting the density profiles to clusters formed hierarchically and non-hierarchically will be given. 

\subsection{Introduction}
\label{Sec.UniveralProfile.Intro}
Over the years, a number of forms for the density profile of a system of collisionless particles has been suggested.  Of particular relevance to the dark matter in galaxy clusters are the Hernquist profile and the Navarro, Frenk, and White (NFW) profile.  Both have a limiting form in the near and far field with a smooth transition about some characteristic radius.  

The Hernquist profile \citep{Hern}, suggested from observations of spherical galaxies, is given by,
\begin{equation}
\frac{\rho(r)}{\rho_c} = \delta \frac{r_s^4}{r(r_s+r)^3},
\label{eq.Prof.Den.Her}
\end{equation}
characterised by a length scale, $r_s$, and an overdensity, $\delta$.

The work of \citet{NFW95,NFW96a,NFW96b} has built a case for a universal density profile of the similar form,
\begin{equation}
\frac{\rho(r)}{\rho_c} = \delta \frac{r_s^3}{r(r_s+r)^2}.
\label{eq.Prof.Den.NFW}
\end{equation}
\citet{SW} (hereafter SW) claims that this form is a byproduct of hierarchical structure formation.  It gives the generalised density profile form,
\begin{equation}
\frac{\rho(r)}{\rho_c} = \delta \frac{r_s^\beta}{r^\alpha(r_s+r)^{\beta-\alpha}}.
\label{eq.Prof.Den.SW}
\end{equation}
The exponents, $\alpha$ and $\beta$, correspond to the exponential dependence of the density on radius in the near and far field.  That is, for $r\ll r_s, \rho \propto r^{-\alpha}$ and equivalently, for $r\gg r_s, \rho \propto r^{-\beta}$.

The forms of Hernquist and NFW are particular cases of the SW density profile. For the Hernquist profile, $\alpha = 1$ and $\beta = 4$.  Correspondingly for the NFW profile, $\alpha = 1$ and $\beta = 3$. It is the shape of the density profile in the inner radii which determines $\alpha$.  The inner radii, however, are affected strongly by the resolution of the simulation.  Indeed, the cluster profiles generally span only two orders of magnitude in radius, the bulk in the outer halo.  Any attempt to fit a smoothly varying curve over this span and then glean information about the near-field and far-field dependency of the profile is thus problematic.

In order to circumvent the ambiguity in derived values of $\alpha$ and $\beta$ due to the limited range of the fit, a non-smoothly varying form is required which is discontinuous in the first derivative at the `knee' separating the nominal near and far fields. This suggests the following functional form should be fit,
\begin{equation}
\frac{\rho(r)}{\rho_c} = \cases{
 \delta_\alpha r^{-\alpha}, & $r < r_s$; \cr
 \delta_\beta  r^{-\beta},  & $r > r_s$;
}
\label{eq.Prof.Den.suggested}
\end{equation}
Since the profile must be continuous at $r_s$,
\begin{equation}
\delta_\beta=\frac{r_s^\beta}{r_s^\alpha}\delta_\alpha.
\label{eq.Prof.Den.suggestedII}
\end{equation}

Simulations of hierarchically formed clusters agree that the NFW profile is an adequate description of the dark matter density distribution.  \citet{CL96} concludes this using a series of simulations of $\Omega = 1$ universes in which the initial density perturbation power spectrum is varied.  Modelling clusters in a variety of cosmologies at very high resolution ($256^3$ particles) \citet{Thomas98} found the density profiles of the clusters to follow the form of the NFW profile.  These simulations were N-body only. 

Similar to the situation being tested here, \citet{HJS98} describes a set of pure N-body simulations of an isolated cluster in which the amount of substructure is controlled via manipulation of the velocity dispersion of the particles.  From this data set, they conclude that the NFW form maintains its universality in non-hierarchical scenarios.

There is not as much agreement as to the details of the density profile. There are several predictions for the shape of the density cusp. They usually involve discussions of the growth of structure from a self-similar initial density distribution. An exception to this is the result of \citet{EC97}. Looking at the stability of clusters to perturbations due to binary encounters, it shows that $\alpha = \fourthirds$ is a stable solution of both the Fokker-Planck and collisionless Boltzmann equations. 

\citet{HS85} assumes a spherically symmetric halo collapse from an initial density perturbation with a power-law exponent of $n$ to derive a radial dependency for the dark matter density of the form
\begin{equation}
\rho(r) \propto r^{-\frac{3(n+3)}{(n+4)}}
\label{eq.Prof.Den.HS}
\end{equation}
for the inner cusp and $\rho(r) \propto r^{-4}$ in the outer halo.  A similar result is found using a more detailed analysis in \citet{Pad96b}. Using the assumptions of self-similarity and stability in the form of the cluster in the non-linear regime (that is, the morphology of a virialised object only scales with time), \citet{Pad96b} derives a relation for the 2-point correlation function in the non-linear stage of evolution from an initial linear density perturbation with a spectral index of $n$,
\begin{equation}
\xi(r) \propto r^{-\frac{3(n+3)}{(n+5)}}.
\label{eq.Prof.Den.Pad}
\end{equation}
Since the 2-point correlation function is simply the excess probability of finding any particle at a distance $r$ from a given particle, if most of the matter is already located in the high-density regions, it is approximately related to the density profile by
\begin{equation}
\frac{\rho(r)}{\rho_{bg}} \simeq 1+\xi(r).
\label{eq.Prof.Den.XiRho}
\end{equation}
This predicts
\begin{equation}
\alpha=3\left(\frac{n+3}{n+5}\right)
\label{eq.Prof.Den.alpha}
\end{equation}
for our density profiles.  The same result is derived in a slightly different form in \citet{Pad96a}.  Linear theory for the growth of clusters from an initial density fluctuation power spectrum, $P(k) \propto k^n$ predicts the same expression for $\alpha$ for $\rho \propto r^\alpha$ in the case of $r\ll r_s$.

Extending the previous arguments to hierarchical clustering, \citet{SW} argues
that the initial cusp form found in the first objects formed will be maintained despite mergers.  Consider a cluster absorbing a smaller satellite.  If the density cusp in the cluster is steeper than that of the satellite, then the satellite will be tidally destroyed, softening the cluster cusp.  If the satellite cusp is steeper, it will survive tidal disruption and sink to the bottom of the cluster, steepening the cluster cusp.

Based on the self-similar infall model, which excludes hierarchical clustering, \citet{HW98} finds $\rho \propto r^{-2}$ for the inner dark matter density profile and $r^{-3}$ in the outer limits.

Observational support exists for the NFW profile.  \citet{Carlberg97} claims agreement of the NFW profile with the average mass-density profile measured from 15 clusters in the Canadian Network for Observational Cosmology (CNOC) cluster survey.  It finds $0.2 \le r_s \le 0.3$ for the average profile of their sample. However, it notes that the fit to a Hernquist profile has smaller residual errors.  Others disagree. \citet{Adami98} finds that the King profile, first suggested for globular clusters \citep{King62}, which has a core, fits the majority of a sample of clusters extracted from the ENACS (ESO Nearby Abell Cluster Survey) better than the NFW form.  It notes, however, that this is not the case for every cluster in the sample; some are consistent with an NFW profile.

\subsection{The density profiles of the simulated clusters}
\label{Sec.Profiles.ResultsDM}
Dark matter density profiles were found for all the clusters formed hierarchically.  These profiles are scaled readily by $R_{200}$ for each cluster to agree to a common form (\fig{fig.Res.Prof.allDM}). It is clear that there are three regimes, nominally the near-field, the far-field, and the background, each with distinctly different power law dependencies.  Only the near and far fields concern us, here.

The four forms of density profiles, NFW, Hernquist, SW, and the form suggested here (\eq{eq.Prof.Den.suggested}), were fit to the scaled SPH-estimated densities of the particles for the unsmoothed hierarchical run (\fig{fig.Res.Prof.Form.fits.hier}).
The method of fitting minimises $\chi^2$.  To separate the mean background density from the fitting, only the span $r<2 R_{200}$ was fit.  The parameters for the fits are outlined in \tab{tab.Res.Prof.Form.coef.101}.

All forms fit the data equally well.  However, due to the limited range in radii over which to fit the profiles (typically only two orders of magnitude), the limiting cases of the near and far field are poorly discerned.  In particular, the value for $\alpha$ is poorly constrained. It is apparent that the `knee' position, $r_s$, and the near-field dependency, $\alpha$, are being used as free parameters to shape the smooth transition between the two regimes.  Consequently, neither represent the transition radius and near-field dependency one would expect. The SW fit fails entirely to discern an inner slope.  However, for the outer slope, the SW fit, as well as the form suggested here, agree with the expectation of NFW.

For these simulations, $n=-1$ which gives $\alpha = 1.5$ using the predictions of \eq{eq.Prof.Den.alpha}. This compares quite well with the results of the fit to the form suggested.  It also agrees with the results of fitting to the form suggested by Syer and White to the non-hierarchical data, for which the near-field is properly fit since the characteristic radius is further out.  However, there is no reason that \eq{eq.Prof.Den.alpha} should hold for non-hierarchical growth.

Other groups find similar results.  For a Cold Dark Matter (CDM) initial power spectrum, \citet{Moore97} and \citet{Moore98} find $\alpha = 1.4$ in a high-resolution (softening of $5 \kpc$) collisionless simulation. More generally, \citet{FM97} finds for the core cusp $\alpha = 1$ to $2$, in high-resolution collisionless simulations of a CDM model using the {\sc GRAPE} special-purpose N-body hardware.  The authors reported a resolution of $1 \kpc$ in these simulations.

The results of the fits for all the models (\tab{tab.Res.Prof.Form.coef.all}) indicate that the fits to the NFW form do as well as the less constrained form suggested here, even in the non-hierarchical cases. The scale radius increases with increased smoothing for the NFW fit in the high resolution simulation, but not the low resolution one.  The inner profiles, as parametrized by $\alpha$, have no dependency on the degree of smoothing but do exhibit a great deal of scatter.  The mean value for the high resolution runs is $\mean{\alpha}=1.8\pm 0.2$.  The outer profile does tend to become less shallow as the smoothing increases.  This is qualitatively consistent with \citet{CER94} who find the profile becomes more shallow as the initial power spectrum index, $n$, is increased in a set of pure N-body simulations of clusters.

\renewcommand{\thisdir}{M-T-scaling}
\section{MASS-TEMPERATURE SCALING LAW}
\label{sec.MT}
In a hierarchical clustering scenario, the process of merging `recreates' the cluster morphology on a continual basis.  Clusters at a given epoch are being formed from structures created at a variety of earlier ages which are, in turn, formed by the amalgamation of many other smaller structures. In a non-hierarchical clustering scenario, the matter in the clusters has actually collapsed for the first time only recently. In both cases, if hydrostatic equilibrium has been established and there exist density and temperature profiles common to all clusters, then it can be expected that there will be a common mass-temperature scaling law for the clusters in a given clustering scenario.

This universality of the gas density and temperature profiles has not, however, been established.  It will be shown that the profiles affect the normalisation, $\tau$, of the mass-temperature relationship.  Presuming the expression, $T=\tau M^{\twothirds}$, the steeper the profiles are, the lower the normalisation, $\tau$, is.

In what follows, we will derive the mass-temperature scaling relation, making our assumptions clear.  The normalisation factor will be more precisely defined, including its dependence on the shape of the gas density and temperature profiles which will then be examined.  The actual normalisation factor for the simulated clusters will be presented, followed by a comparison with that determined implicitly from the derived relation combined with the information of the profile shapes. 

\subsection{Derivation of the mass-temperature scaling law}
\label{sec.MT.derivation}
The equation of hydrostatic equilibrium relates the pressure, $P(r)$, with the  the mass, $M(r)$, internal to the radius, $r$, via
\begin{equation}
\frac{GM(r)}{r^2} = -\frac{1}{\rho}\frac{dP}{dr},
\label{eq.MT.hydro_basic}
\end{equation}
where $\rho$ is the gas density and $G$ retains its normal use as the gravitational constant. Combined with the perfect gas law,
\begin{equation}
P = \frac{\boltz \rho T}{\mu m_u},
\label{eq.MT.perfect_gas}
\end{equation}
which relates the pressure to the temperature, $T$, with $\boltz$, $\mu$, and $m_u$ being Boltzmann's constant, the mean molecular weight, and the atomic mass unit, respectively, \eq{eq.MT.hydro_basic} gives,
\begin{equation}
\frac{GM(r)}{r} = -\frac{k T}{\mu m_u} \left( \frac{r}{\rho}\frac{d\rho}{dr} + \frac{r}{T}\frac{dT}{dr} \right).
\label{eq.MT.hydro_inter}
\end{equation}
Using the definition of $\ln$,
the following relation is found:
\begin{equation}
\frac{GM(r)}{r} = -\frac{kT(r)}{\mu m_u}\left[ \frac{d\ln \rho}{d\ln r}+\frac{d\ln T}{d\ln r} \right].
\label{eq.MT.hydro}
\end{equation}

It is shown in \sects{sec.MT.GasDensityProfile}~and~\ref{sec.MT.TauProfile} that both the gas density and temperature profiles can be approximated by power laws away from the `knee' radius.  Taking the exponents of these respective profiles to be $n_\rho$ and $n_T$ gives,
\begin{equation}
\rho(r) \propto r^{n_\rho}, T(r) \propto r^{n_T}\\
\Rightarrow \frac{GM(r)}{r} = -\frac{kT(r)}{\mu m_u}\left[ n_\rho + n_T \right].
\label{eq.MT.hydro2}
\end{equation}
In general, $n_\rho$ and $n_T$ are negative.

The properties of clusters are characterised by a radius, $R_{\bar{\delta}}$, and mass,  $M_{\bar{\delta}}$, of some overdensity.  This overdensity is typically taken to be around the value of the virial overdensity for a top-hat collapse, i.e. $\bar{\delta}=178$.  However, the characteristic radius is not particularly sensitive to the choice of $\bar{\delta}$.  Given this description, we have the definition
\begin{equation}
R_{\bar{\delta}} = \left[ \frac{3}{4\pi} \frac{M_{\bar{\delta}}}{\bar{\delta}\rho_c} \right]^{1/3}
\label{eq.MT.Rcrit}
\end{equation}
which, when combined with \eq{eq.MT.hydro2}, gives
\begin{equation}
\G \left[ \frac{4\pi}{3} \bar{\delta}\rho_c \right]^{1/3} M_{\bar{\delta}}^{2/3} = -\frac{kT(R_{\bar{\delta}})}{\mu m_u}\left[n_\rho + n_T \right].
\label{eq.MT.TM}
\end{equation}
Thus the scaling relationship, $T \propto M^{2/3}$, is found.  Recall that the critical density is given by,
\begin{equation}
\rho_c = \frac{3 (h 100\km \s^{-1} \Mpc^{-1})^2}{8\pi \G}.
\label{eq.MT.rhoc}
\end{equation}
On the scales of interest, this relation may be written as,
\begin{equation}
\frac{T(R_{\bar{\delta}})}{1 \keV} = \frac{2.83 h^{2/3} \bar{\delta}^{1/3}}{-(n_\rho + n_T)} \left[ \frac{M_{\bar{\delta}}}{10^{15}\Msun} \right] ^{\frac{2}{3}} .
\label{eq.MT.TM2}
\end{equation}

Significantly, the coefficient of the mass-temperature relation is dependent on the forms of the gas density and temperature profiles through the exponents $n_\rho$ and $n_T$.

\subsection{The temperature normalisation parameter: $\tau$}
\label{Sec.MT_law.Tau}
Since the temperatures of the clusters are scaled by the mass of the cluster, it is convenient to introduce the parameter
\begin{equation}
\tau(r) \equiv \frac{\frac{T(r)}{1 \keV}}{\left( \frac{M_{200}}{10^{15}\Msun} \right)^{\frac{2}{3}}}.
\label{eq.Prof.Temp.tau}
\end{equation}
Thus, the $\tau$ profile for a cluster will have the same form as $T(r)$ but will be scaled by the cluster mass.  Hence, if $T(r) \propto r^{n_T}$, then $\tau(r) \propto r^{n_T}$ as well.

For ease of comparison, two temperature parameters are defined.  The first, 
\begin{equation}
\tau_{num} \equiv \frac{\frac{T(r=0)}{1 \keV}}{\left( \frac{M_{200}}{10^{15}\Msun} \right)^{\frac{2}{3}}},
\label{eq.Prof.Temp.tau_num}
\end{equation}
can have its value found from the distribution of temperature versus mass for the clusters.
 
Given that \eq{eq.MT.TM2} gives $T(R_{\bar{\delta}})$ while $\tau_{num}$ is related to $T(r=0)$, we can compare the two by including a factor for the scaling between $r=R_{\bar{\delta}}$ and $r=R_{iso}$, the radial limit of the isothermal core.  This gives for the second temperature parameter,
\begin{equation}
\tau_{analy} = \frac{2.83 h^{2/3} \bar{\delta}^{1/3}}{-(n_\rho + n_T)} R_{iso}^{n_T},
\label{eq.Prof.Temp.tau_analy}
\end{equation}
which will allow us to compare the actual central temperatures with the central temperature to be expected for a cluster with known density and temperature radial profiles at $R_{\bar\delta}$ which have power-law dependencies of $n_\rho$ and $n_T$ respectively as well as isothermal radii approximated by $R_{iso}$. Both of these normalisation parameters may be measured directly and independently from the simulated data.  Their equivalence would verify the derived M-T scaling law encompasses the relevant physics.

\subsection{Mass-temperature distribution}
\label{sec.MT.Sims}
The mass-temperature relation was derived directly from the simulated data using the mean temperature in the core of the clusters and the mass within the virial radius, $R_{200}$. Temperatures within $0.2 R_{200}$ are essentially constant (see \sect{sec.MT.TauProfile}). This distance was used as the cutoff radius in determining the mean cluster temperature, $T_r$.  The distribution of temperature versus mass is illustrated in \fig{fig.MT.TvsM}.

Fitting with a free coefficient for the relation $T\propto M^n$, the correlation between central cluster temperature, $T_r$, and the mass of the cluster is found to obey the relationship,  
\begin{equation}
T_r = \tau \left( \frac{M_{200}}{10^{15}\Msun} \right)^{2/3} \keV,
\label{eq.Res.MT.Tr}
\end{equation}
since $n=0.67\pm 0.04$, invariant of the degree of hierarchical clustering. The values for $n$ for each simulation are listed in \tab{tab.Res.MT.freecoef}.

For \tab{tab.Res.MT.freecoef}, errors were estimated using a bootstrap approach. Fits were calculated repeatedly using cluster data from a sample of half of the clusters, chosen randomly.  The errors were determined from the variance of the coefficients of these fits.  This method, of course, biases towards the more common small-mass clusters whose counterparts in the real world would be less readily observed.  When the samples are restricted to the more massive clusters, the results do not change significantly. Fits to the data points assume a correlation among clusters and, as such, lead to smaller errors than are found assuming independence among clusters (cf. \eq{eq.Res.MT.error}).

Using a fixed value of $n=\twothirds$, the normalisation factor, $\tau_{num}$, was calculated for the clusters within each simulation (\tab{tab.Res.MT.coef}).
The data points were assumed to be independent for the sake of the fit and error estimate.  That is, the error is,
\begin{equation}
\sigma = std \left[ \frac{T_r/\keV}{\left( \frac{M_{200}}{10^{15}\Msun} \right)^{2/3} } \right] .
\label{eq.Res.MT.error}
\end{equation}
There may be a weak ($1\sigma$) dependence of $\tau_{num}$ on the degree of smoothing in the initial conditions.  The value of $\tau_{num}$ drops from $6.8\pm 1.6$ in the unsmoothed case to $5\pm 1$ for the $7 \Mpc$ smoothing.  This is consistent with the trend in $\tau_o$ found for the fits to $\tau$-profile. 


The value for the normalisation coefficient, $\tau_{num}$, is in agreement with the number of numerical simulations, albeit on the high side. For the simulations of six clusters formed in a cold dark matter (CDM) model described in \citet{NFW95}, a normalisation of $5.3\pm 0.7$ was found.  Simulations of a variety of cosmological models which varied in $\Omega_o$ are described in \citet{EMN}.  There, a factor value of $5.20$ is reported for $\tau$.  Note that for this last value, the empirical translation $M_{200} \approx 1.2M_{500}$ has been assumed here. From the data presented here, this ratio is found to be $1.17 \pm 0.11$ and a slight trend towards higher values for more massive clusters is noted.  For the previous assumption, this trend is insignificant.  For the ten clusters described in \citet{ENF98}, a normalisation of $6\pm 1$ fits their data after scaling from the overdensity of 100 used in their analysis to the overdensity of 200 used here.  These clusters were simulated using initial conditions for a low density, flat universe. Using an Eulerian code, \citet{BN98} found a lower value of $4.7\pm0.1$ from a variety of CDM models.  \citet{BBP98} has shown that if the gas is preheated and allowed to collapse adiabatically into an isothermal potential well, then the mass-temperature scaling relationship overestimates the halo masses by up to an order of magnitude.  This occurs for haloes the size of groups of galaxies or less ($M < 10^{14}\Msun$), which is at the low end of the size of the haloes examined here.  Otherwise, it recovers the relationship found here, with a coefficient of about $4.5$.

The weighted mean of the value for $tau$ using $\tau_{num}$ and the values mentioned above is $5.1 \pm 0.1$.  For this, an error of 5 times the least significant digit was attributed if none was otherwise given.

The effect of galactic winds on the state of clusters is addressed in \citet{ME97}.  It reports that the gas density profiles are made more shallow by the winds while the central temperatures are not.  The gas density profile parameter, $n_\rho$ rises from $-2.34$ to $-1.75$ with the inclusion of winds. As for the mass-temperature relation, for their sample of 18 clusters, the normalisation factor changes slightly from $4.8\pm 0.4$ to $5.2\pm 0.3$ with the inclusion of galactic winds.  That the factor rises as the density profile becomes more shallow is consistent with \eq{eq.Prof.Temp.tau_analy} in trend, if not magnitude, since a value of $4.8$ would rise to $5.5$ if $n_T$ were held constant at $-0.5$. \citet{ME97} does not report a value for $n_T$ though it indicates that the winds lead to a steeper temperature profile.  This would partially offset the increase in $\tau$ due to the decrease in $n_\rho$.

Various authors have found observational evidence supporting this scaling law. \citet{Schind96} and \citet{TKB} find $T=7.8 \keV (M_{Tot}/10^{15} \Msun)$, with the cosmological factors, $h$, set to that used in these studies ($0.65$).  \citet{HMS99} verifies the form of the mass-temperature scaling relation using a sample of 23 clusters extracted from catalogues of clusters for which temperatures have been determined using {\em ASCA} \citep{Fukazawa97, Markevitch98} and masses via the projected velocity dispersion of member galaxies \citep{Girardi98}.  It reports a normalisation factor of $6.4$.

\subsection{Mean gas density profiles}
\label{sec.MT.GasDensityProfile}
In order to calculate a value for $\tau_{analy}$, the gas density profile is required.  A mean profile is determined for the clusters within a simulation using the gas density versus radius distribution for the cluster gas particles.  The radii are scaled by $R_{200}$ under the assumption that the gas will primarily follow the dark matter whose density distribution has been seen to scale with the virial radius, $R_{200}$.

The gas densities were recalculated using the SPH density estimator with the smoothing lengths set to enclose exactly $N_{SPH}=32$ particles.  The distance from the cluster centres, scaled by the overdensity radius $R_{200}$ for each of the clusters, was found for each particle.  The sample of particles comprised those that are within $2R_{200}$ but beyond the point interior to which there were 300 particles. This interior limit removes those particles for which the SPH density estimates are erroneous due to the steep density gradient (see \app{sec.Tech.Grad}).  \fig{fig.MT.GasDensity.101} illustrates the sample for the clusters formed hierarchically.  The cluster profiles are scaled remarkably well by the virial radius, $R_{200}$, considering that this is a combination of gas particles from 100 clusters spanning a range in mass of 30 times. It was verified that this scaling by the virial radius holds for the non-hierarchical simulations, also.  A density profile was fit using the discontinuous density profile form given by \eq{eq.Prof.Den.suggested} introduced in \sect{Sec.Profiles.ResultsDM}.  The results are summarised in \tab{tab.Res.Prof.Gas.Rho}.

There is a weak trend in which $r_s$ increases with increased smoothing.  A much more significant trend is for $\alpha$ to decrease with increased smoothing.  That is to say the density profile becomes more shallow as smoothing is increased.  The outer profile changes little with a power-law dependency on radius of $\approx -2.7$ which is slightly shallower than the density profile for the dark matter (see \sect{Sec.Profiles.ResultsDM}).

The standard form to which gas density profiles are fit is the ``$\beta$''-model \citep{CF76} which describes the expected density profile of an isothermal cloud. This model is used frequently in the interpretation of x-ray observations of galaxy clusters \citep{FG83,JF84}.  It has the continuous form
\begin{equation}
\rho(r) = \rho_o\left[1+\left(\frac{r}{r_s}\right)^2\right]^{-\frac{3}{2}\beta}.
\label{eq.MT.GasDensityProfile.betaprofile}
\end{equation}
The significance of $\beta$ in this fit is that it also represents the ratio of the galaxy (or more specifically, collisionless baryonic component) kinetic energy to gas thermal energy.  That is,
\begin{equation}
\beta \equiv \frac{\sigma_{gal}^2}{\boltz T / \mu m_u} ,
\label{eq.MT.GasDensityProfile.beta}
\end{equation}
where $\sigma_{gal}$ is the one dimensional velocity dispersion for the galaxies in the cluster.  Though there is no collisionless baryonic component in these simulations and the gas is well thermalized, it may still be interesting to see if the inferred value of $\beta$ varies among the models.  When \eq{eq.MT.GasDensityProfile.betaprofile} is fit to the mean gas profiles (\tab{tab.Res.Prof.Gas.Rho.Beta}), it is found that $\beta$ varies little among the models with a mean value of $0.82\pm0.03$ for the high resolution runs.  This compares with the value of $0.76\pm0.06$ found by \citet{TKB} who analysed a single cluster simulated at high resolution in a CDM model.  \citet{ENF98} simulated clusters in a low-density universe and found a mean value of $0.74\pm0.15$.

\subsection{Mean $\tau$ profile}
\label{sec.MT.TauProfile}
Like the gas density profile, the mean temperature profile is required to calculate $\tau_{analy}$. Since the cluster temperature is scaled by the mass, it can be expected that the temperature profiles are also accordingly scaled. As such, the profile of the scaled-temperature parameter, $\tau$, defined in \eq{eq.Prof.Temp.tau}, should be used to compare temperature distributions.

A data set for the temperature parameter, $\tau$, was created in a manner similar to that of the gas density data set in \sect{sec.MT.GasDensityProfile}.  For the hierarchical case, the mean scaled $\tau$~profile is shown in \fig{fig.MT.GasTau.101}. There is a great deal more scatter than for the density.  Some of this is due to smaller clusters being satellites of larger clusters.  The halo gas of these smaller clusters gets shocked to high temperatures which then are scaled by the mass of the smaller cluster. However, there is an approximately isothermal core that extends to $\approx 0.2 R_{200}$. The outer profile may be crudely fit with a power-law dependency on radius.  Fits to the profiles with the discontinuous form given in \eq{eq.Prof.Den.suggested} which has a free power-law index for the inner regime, $\alpha$, find values of $\alpha$ spanning $-0.05$ to $0.10$. For this reason, an approximate $\tau$-profile is fit using a form similar to the discontinuous form but with an iso-thermal core:
\begin{equation}
\tau(r) = \cases{
 \tau_o, & $\frac{r}{R_{200}} < R_{iso}$; \cr
 \tau_o \left( \frac{R_{200}}{R_{iso}} \right)^{n_{T}} \left( \frac{r}{R_{200}} \right)^{n_T}, & $\frac{r}{R_{200}}  > R_{iso}$;
}
\label{eq.Prof.T.suggested}
\end{equation}
where $\tau_o$ is the scaled temperature (in units of $T/M^{2/3}$) in the centre of the cluster.  The results of fitting this to the scaled temperature versus scaled radius distributions for each of the simulations are given in \tab{tab.Res.Prof.Gas.T}.

The central temperatures are found to generally decrease with increased smoothing. The isothermal core radii increase slightly with increasing smoothing, in step with the trend found for $r_s$ of the gas-density profiles, but marginally interior to $r_s$.  The power-law dependency on radius, $n_T$, shows a trend to steepen slightly with increased smoothing, ranging from -0.4 to -0.6.  This value has a strong dependency on resolution, with $n_T \approx -0.8$ for the lower resolution runs, as does the iso-thermal radius which increases in diameter from $0.1$ to $0.5 \fraction{r}{R_{200}}$ as the resolution is halved.  This is consistent with the two-body heating problem described in \sect{sec.Sims.2bodyHeating}. It was determined, there, that two-body interactions may be able to heat the gas to the temperature of the core over radii of $0.1$ to $0.2\Mpc$ for the high resolution run and $0.1$ to $0.5\Mpc$ for the lower.  As such, the scaled temperature profile may become even more shallow with a further increase in resolution.

\subsection{Comparison with the semi-analytic prediction}
\label{sec.MT.Pred}
Recall \eq{eq.Prof.Temp.tau_analy}, which defines,
\begin{equation}
\tau_{analy} \equiv \frac{2.83 h^{2/3} \bar{\delta}^{1/3}}{-(n_\rho + n_T)} R_{iso}^{n_T}.
\label{eq.Res.Tau.analytic}
\end{equation}
We can compare $\tau_{analy}$ to that inferred from the numerical simulations using $n_\rho = \beta$ from \tab{tab.Res.Prof.Gas.Rho} and $n_T$ from \tab{tab.Res.Prof.Gas.T}. \tab{tab.Res.Tau.coef} summarises the results with the values of $\tau_{analy}$ given in the second column.

Within error, the values for $\tau$ from the scaling law fit, $\tau_{num}$, the $\tau$-profile fit, $\tau_o$, and the semi-analytic expectation assuming hydrostatic equilibrium, $\tau_{analy}$, agree with the exception of the values for the $7 \Mpc$ top-hat model.  However, the trend for $\tau$ to decrease with increasing smoothing is not as obvious when $\tau_{analy}$ is considered.  In \sects{sec.MT.GasDensityProfile} and~\ref{sec.MT.TauProfile}, it is seen that $n_\rho$ is invariant and $n_T$ is only weakly variant among the runs.  It is the increase in $R_{iso}$ with increased smoothing that leads to the decrease in $\tau_{analy}$ (recalling that $n_T$ is negative).  The slight increase in $n_T$ with smoothing tends to offset this trend.  Consequently, within the scatter among clusters, the value for $\tau$ does not vary appreciably and could be considered a constant among cosmologies.

\section{Conclusions}
\label{sec.Conclusions}
Using a series of numerical simulations, the importance of hierarchical clustering to the state of the matter in cluster of galaxies was examined.  The simulations contained a collisional component, permitting an examination of the hot gaseous halo of the clusters along with the dominant collisionless dark matter component.  The comparison was carried out using simulations in which structure formed hierarchically, from smaller initial structures, and those in which the formation of these earlier structures was suppressed by smoothing the initial density distribution.

The clusters within each simulation were not examined individually, but instead were examined as a whole sample.  As required, individual cluster parameters were scaled by either the radius or mass at an overdensity of 200.

We find the overall morphology of the simulated clusters robust to their formation method and relatively independent of the hierarchical clustering nature of structure formation.  In particular, the dark matter profile and the mass-temperature relationship do not vary substantially between clusters formed hierarchically and those formed otherwise.  However, these features do vary systematically in small ways which may be significant to both theory and observations.  These will be discussed in the following sections.

It is noted that the gaseous component of the simulated clusters for this work were shown to be in hydrostatic equilibrium to beyond their respective virial radii, as indicated by $R_{200}$.  At this radius, the ratio of the actual gas pressure to the gas pressure expected, from the integration of the equation of hydrostatic equilibrium, is within $20\%$ of unity for the majority of the clusters.

\subsection{Universal density profile}
\label{sec.Conclusions.UnivProf}
The density profile form of \citeauthor{NFW95} is found to fit the mean dark matter density profiles of the clusters of all models studied, indicating that {\em the universal profile is not a by-product of hierarchical clustering}.
This profile form, however, fits only marginally better than the Hernquist profile.  Indeed, the discontinuous form suggested here generally fits as well, if not better, than the NFW form.  However, it has more free parameters.

The discontinuous form indicates that the inner regimes of the dark matter density profile are approximately dependent on radius with $\rho\propto r^{-1.8}$, independent of the degree of smoothing of the initial conditions.  The outer regimes, in contrast, do seem to be dependent on smoothing.  The profile becomes marginally shallower as the degree of smoothing is increased.  The dependency varies from $\rho\propto r^{-2.7}$ in the case of the hierarchical, unsmoothed case to $\rho\propto r^{-2.3}$ for the 7 Mpc top-hat smoothed model and the $(14 \Mpc)^{-1}$ frequency cutoff model.  The characteristic radius, $r_s$, is affected by the smoothing; it expands from $0.1 R_{200}$ to $0.2 R_{200}$ when hierarchical clustering is eliminated.  This is consistent with the later epoch of formation of the cluster.

The NFW profile has been established to be a robust description of the dark matter density profile, independent of hierarchical clustering and, by extension, independent of the assumed cosmology.  This strongly implies that dark matter in galaxy clusters is distributed in this manner.  If detailed mass maps of clusters using, for example, gravitational lensing observations indicate that the mass is not distributed in this fashion, it would indicate that either numerical simulations are missing some significant physics or the Standard Model is flawed in a fundamental manner.  It will be difficult to attribute the discrepancy to the assumed initial density perturbation spectrum within the Standard Model.  Unfortunately, this also implies that the observed mass density profiles of clusters will tell us little about this initial perturbation spectrum.

\subsection{Mass-temperature scaling law}
\label{sec.Conclusions.MT}
The clusters were found to have iso-thermal cores.  Using the temperatures within these cores and the masses within an overdensity radius of 200, a normalisation factor, $\tau_{num}$, for each simulation was found. Integrating the equation of hydrostatic equilibrium permitted a derivation of the temperature parameter, $\tau_{analy}$, which is dependent on the radial profiles of both the gas density and the temperature as well as the radius of the iso-thermal core.  Mean profiles of these parameters were determined for the clusters within each simulation. This permitted the determination of values for $\tau_{analy}$ which were compared with the respective values of $\tau_{num}$ to both verify the model and explain any variation found between the models.

While determining gas density profiles for calculating $\tau_{analy}$, the value of $\beta$ was also determined.  This factor is related to the ratio of galaxy kinetic energy to baryonic thermal energy in real systems as well as being the $\beta$ used in fits of the gas distribution in the data from observations of clusters.  It is found to not vary with the degree of smoothing. Instead, it maintains a value of $0.82\pm0.03$.

The mass-temperature scaling law (\eq{eq.Res.MT.Tr}) is dependent on the degree of hierarchical cluster formation, albeit at a $1\sigma$ level.  The temperature parameter $\tau_{num}$ varies from $7\pm 1$ in the unsmoothed case to $5 \pm 1$ in case of the $7\Mpc$ top-hat run with the values for the intermediate smoothing models lying in between.  The trend is consistent with the expectation of hydrostatic equilibrium.  This value decreases from $7.7\pm 0.2$ to $6.4\pm 0.7$ over the span of smoothing.  There is a systematic discrepancy between the expectation, $\tau_{analy}$, and the value derived directly from the clusters, $\tau_{num}$, with the semi-analytic expectation consistently larger than the measured value. This may be attributed to numerical, two-body heating which is found to be significant in the domain of $R_{iso}$. But again, the discrepancy is less than the cluster-to-cluster variation within a simulation. The results of the fits to the profiles indicates that it is the variation in the isothermal radius that explains the variation with smoothing.  The isothermal radius increases as the degree of hierarchical clustering is reduced.  This occurs in step with the gas density profiles which become shallower in the inner radii.

\bibliographystyle{plainnat_mnras}
\bibliography{biblio}
\renewcommand{\thisdir}{Appendix}
\appendix
\section{SPH NEAR STEEP DENSITY GRADIENTS}
\label{sec.Tech.Grad}
It is known that the SPH estimation of the density for a particle fails near a steep density gradient.  The density is essentially determined by finding the radius of a sphere that encompasses some given number of particles, $N_{SPH}$, which is typically 20-30. That the particle contributions are weighted by a smoothly-varying kernel function is a second-order correction to the estimate.  Crudely, in the presence of a varying density gradient, contributions from particles in the higher-density region exceed those from the lower-density region, leading to an over-estimate of the local density. Non-spherically symmetric kernels alleviate this problem, but are computationally expensive both in memory resources and CPU time. In this appendix, this phenomenon will be explored in more detail with the aim to predict those gradients that suffer this effect most strongly.

For any spherically symmetric weighting kernel, the SPH estimate of the density for a particle in a {\em constant} density gradient will produce no error.  This requires, of course, that the gradient be constant over the span of $x-2h<x<x+2h$ which implies that the gradient is resolved by the particle density.  Here, we will examine the error induced by a varying density gradient.

The toy model that will aid us in our exploration will be a volume with an increasing density gradient in one direction.  Specifically, the density will behave as $\rho \propto x^{-1}$.  This is a form similar to the density profiles found in cosmological objects.  An iso-density surface for these objects, though spherical in morphology, is locally flat.  Hence this toy model is relevant to the density gradients found in cosmological objects.

The models were constructed from 100 slices, each with the same number of particles but with widths, $W_i$, such that $ W_{i-1} F = W_i = W_{i+1}/F $ where $F \geq 1$.  To test variation with form, $F$ is set here to $1.1$ and $1.5$ for two different sets of gradient fields.  The particles were distributed randomly.

In our tests, the mass per particle is unity, giving a number density, $n$, equivalent to the mass density, $\rho$.

Error in the density estimate arises from both statistical variance as well as the systematic error due to the density gradient.  The former error is reduced by using more particles in the estimate of the density.  The latter is reduced by kernel averaging over a smaller region, and hence using fewer particles.  The regime of dominance of these effects was explored by calculating the density using $N_{SPH}=$ 64, 32, 16, and 8 neighbour particles. Near the density cusp, where gradients are higher, the density estimate which averages over fewer particles ($N_{SPH}\le 16$) displays large variance, but no systematic over-estimation of the density.  In comparison, the estimates using larger numbers of particles ($N_{SPH} > 16$) systematically over-estimate the density by a factor of 10 to 20\% within $x=0.1$ of the cusp but exhibit \onethird\ less variance.  Further away from the cusp, where the density gradient is smaller, the larger values of $N_{SPH}$ allow density estimates with smaller amounts of statistical variance and no systematic bias.

Increasing the number density of particles increases the resolution of the simulation.  This permits steeper gradients to be resolved.  For the $\rho\propto x^{-1}$ profile of our tests, this behaviour was qualitatively verified by varying the number of particles per slice. Numbers of 50, 100, 200, 500, and 1000 particles per slice were used.  For the gradient regime, $\frac{dn}{dx} > 500$, the smaller number densities increased the error in the density estimate.  The gradient field that used 1000 particles per slice was able to go to an order of magnitude greater density gradient.

More useful is a parametrization of the density gradient that is independent of the local number density of particles.  Consider the parametrization based on the change of the density over the smoothing distance, $2h$, given by $\rho(x-h)/\rho(x+h)$. It was found that this parametrization is independent of the number of particles per slice.  It is also independent of the form of the gradient. It indicates that SPH fails (at the 20\% level) when the density across a smoothing length varies by more than a factor, $F_c$, or 3-4 times, for the $\rho\propto x^{-1}$ profile used here. 

An estimate of the radial distance from the centre of a dense clump at which SPH will begin to calculate erroneous density values can be made using $F_c$.  This factor, $F$, can be defined as
\begin{equation}
F\equiv \frac{n(r-h)}{n(r+h)}
\label{eq.Grad.Fdef}
\end{equation}
where $n$ is the number density of particles, or $\rho/m$.  The critical radius, $r_c$, below which the density estimates will be in error is related as
\begin{equation}
\frac{n(r_c-h(r_c))}{n(r_c+h(r_c))} = F_c.
\label{eq.Grad.r_c-F_c}
\end{equation}
The estimate $n(r+\delta r) \approx n(r)+\delta r \frac{dn}{dr}$ gives
\begin{equation}
\frac{n(r-h)}{n(r+h)}\simeq\frac{n(r)-h(r)\frac{dn}{dr}}{n(r)+h(r)\frac{dn}{dr}}.
\label{eq.Grad.Fapprox}
\end{equation}
If the density profile is assumed to take the form, $n(r)=n_o r^{-\alpha}$ then $dn/dr=-\alpha n_o r^{-\alpha-1}$ leading to
\begin{equation}
\frac{n(r-h)}{n(r+h)}
 \simeq \frac{n_o r^{-\alpha} + \alpha h(r) n_o r^{-\alpha -1}}
             {n_o r^{-\alpha} - \alpha h(r) n_o r^{-\alpha -1}}
 = \frac{1 + \alpha h(r) r^{-1}}{1 - \alpha h(r) r^{-1}}
 = \frac{r + \alpha h(r)}{r - \alpha h(r)}.
\label{eq.Grad.Freduction}
\end{equation}
Using the relation between $h$, $N_{SPH}$, and $n$,
\begin{equation}
h(r)= \frac{1}{2} \left( \frac{3}{4} \frac{N_{SPH}}{\pi n(r)} \right)^{\onethird},
\label{eq.Grad.h_eq}
\end{equation}
and \eq{eq.Grad.r_c-F_c}, the value for $r_c$ can be derived,
\begin{equation}
r_c^{1-\frac{\alpha}{3}} = \frac{\alpha}{2} \left( \frac{3}{4} \frac{N_{SPH}}{\pi n_o} \right)^{\onethird} \frac{F_c+1}{F_c-1}.
\label{eq.Grad.r_c}
\end{equation}
In these studies, $N_{SPH}=32$ and, as seen, $F_c \approx 3$, giving the approximate equation,
\begin{equation}
r_c^{1-\frac{\alpha}{3}} \simeq \frac{2 \alpha}{n_o^{\onethird}}.
\label{eq.Grad.r_c_approx}
\end{equation}
This allows the derivation of an approximate minimum radius for a given cluster with a compression factor (c.f. overdensity) of $\delta_c$ at the critical radius, $r_c$.  The compression factor is defined as the density at the critical radius compared with the mean density in the volume. That is,
\begin{equation}
\delta_c \equiv \frac{\rho_c}{\mean{\rho}_V} = \frac{n_c}{\mean{n}_V},
\label{eq.Grad.delta}
\end{equation}
where the notation $\mean{A}_V$ refers to the volume-weighted mean of the parameter $A$.  Since $ \mean{n}_V = Res^3$, where $Res$ is the number of nodes per side in the initial density distribution (i.e., $N_{gas} = Res^3$),
\begin{equation}
n_c = \delta_c Res^3.
\label{eq.Grad.n_c}
\end{equation}
Recall that $n(r) = n_o r^{-\alpha}$, implying $n_o = \delta_c Res^3 r_c^\alpha$ giving,
\begin{equation}
r_c \simeq \frac{2\alpha}{\delta_c^{\onethird} Res}.
\label{eq.Grad.r_c_estimate}
\end{equation}
For the maximum overdensity observed, $\delta_c=10^6$, $r_c=40\kpc$ for simulations with $64^3$ particles.  For $\delta_c=10^3$, this increases to $r_c=400\kpc$.  Clearly, $r_c$ is enclosing a constant number of particles.  This minimum number of particle, $N_{min} = \frac{4}{3} \pi r_c^3 \delta_c Res^3$, when $r_c$ from \eq{eq.Grad.r_c_estimate} is substituted, gives,
\begin{equation}
N_{min} \simeq 4 (2 \alpha)^3.
\label{eq.Grad.Nmin}
\end{equation}
This number is the minimum number of gas particles, within $r_c$, required to properly resolve the density gradient given by $\alpha$.  For $\alpha=2$, the required number is $250$ but this number is sensitive to $\alpha$ and drops to $100$ for  $\alpha=1.5$

The number, $N_{min}$, was used to set the minimum cluster size in the analysis as well as limit the regime over which bulk properties were determined; radii enclosing fewer than $N_{min}$ gas particles were ignored.

\renewcommand{\thisdir}{Tables/tables}

\begin{table}
\begin{center}
\begin{tabular}{ll}
\hline
    $H_o$ & $65 \km \s^{-1} \Mpc^{-1}$\\
$\Omega_{Dark Matter}$ & $0.9$\\
$\Omega_{Gas}$ & $0.1$\\
$\Lambda$ & $0$\\
Power law index & $-1$\\
$\sigma_8$ & $.935$\\
\hline
\end{tabular}
\end{center}
\caption[Properties of the assumed cosmology]{
Properties of the assumed cosmology.}
\label{tab.Sim.cosmology}
\end{table}

\begin{table}
\begin{center}
\begin{tabular}{ccccccc}
\hline
                        &\multicolumn{2}{c}{\# of particles}& $m_g$ & $m_{DM}$                 & $r_{smooth}$  &$\epsilon$      \\
                        & gas & dark			    &\multicolumn{2}{c}{$10^{10}\Msun$}&($h^{-1} \Mpc$)&($h^{-1} \kpc$) \\
\hline
unsmoothed 		&$64^3$& $64^3$ 		    &  1.03 & 9.27                     & 0             & 20\\
$3\Mpc$ top-hat		&$64^3$& $64^3$ 		    &  1.03 & 9.27                     & 3             & 20\\
$(7\Mpc)^{-1}$ k-cutoff	&$64^3$& $64^3$ 		    &  1.03 & 9.27                     & 7             & 20\\
$7\Mpc$ top-hat		&$64^3$& $64^3$ 		    &  1.03 & 9.27                     & 7             & 20\\
$(14\Mpc)^{-1}$ k-cutoff&$64^3$& $64^3$ 		    &  1.03 & 9.27                     & 14            & 20\\
low-res. unsmoothed     &$32^3$& $32^3$  		    &  8.24 &74.16                     & 0             & 40\\
low-res. $7\Mpc$ top-hat&$32^3$& $32^3$  		    &  8.24 &74.16                     & 7             & 40\\
\hline
\end{tabular}
\end{center}
\caption[Properties of the simulations]{
Properties of the simulations. Given are the number of each type of particle, the mass per particle for both the gas, $m_g$, and dark matter,$m_{DM}$, the effective smoothing radius, $r_{smooth}$, and the gravitational softening length, $\epsilon$. The length of a side of the simulation volume is $40 h^{-1} \Mpc$ for all simulations.
}
\label{tab.Sim.simulations}
\end{table}

\begin{table}
\begin{center}
\begin{tabular}{lccc}
\hline
                        & $N_{clusters}$& $M_{min}[10^{15}\Msun]$  	& $M_{max} [10^{15}\Msun]$ \\
unsmoothed	        & 100		&  0.05				& 1.52 \\
$3\Mpc$ top-hat         & 48 		&  0.05			    	& 2.85 \\
$(7\Mpc)^{-1}$ k-cutoff & 48 		&  0.05				& 1.75 \\
$7\Mpc$ top-hat         & 8  		&  0.03			   	& 1.95 \\
$(14\Mpc)^{-1}$ k-cutoff& 6 		&  0.15				& 2.27 \\
low-res. unsmoothed     & 3 		&  0.54				& 0.70 \\
low-res. $7\Mpc$ top-hat& 5  		&  0.40				& 1.88 \\
\hline
\end{tabular}
\end{center}
\caption[Results of the cluster search]{
Results of the cluster search.  Given for each run is the number of clusters found, $N_{clusters}$, and the range of masses of the clusters, $M_{min}$ and $M_{max}$.}
\label{tab.analysis.clusters}
\end{table}

\begin{table}
\begin{center}
\begin{tabular}{ccccc}
         & $r_s$          & $\alpha$       & $\beta$        & $\chi^2/N$ \\
\hline
NFW      & $0.06\pm 0.01$ & $1$            & $3$            & $0.98$     \\ 
Hernquist& $0.19\pm 0.01$ & $1$            & $4$            & $1.00$     \\ 
SW       & $0.0008\pm 0.003$ & $-7\pm3$    & $2.94\pm 0.02$ & $0.98$     \\ 
here     & $0.12\pm 0.01$ & $1.84\pm 0.03$ & $2.74\pm 0.01$ & $0.97$     \\ 
\hline
\end{tabular}
\end{center}
\caption[Fits to the dark matter density profiles]{
Coefficients for fits to the dark matter density profiles for the hierarchical case using a variety of profile forms.  The SW fit, though free to fit an inner slope, fails to converge to a value for $\alpha$ and fits, instead, a single power law.
}
\label{tab.Res.Prof.Form.coef.101}
\end{table}

\begin{table}
\begin{center}
\begin{tabular}{cccccc}
Run & Form & $r_s$          & $\alpha$       & $\beta$ &      $\chi^2/N$ \\
\hline
\multirow{2}{25mm}{unsmoothed} 
& NFW      & $0.06\pm 0.01$ & $1$            & $3$            & $0.98$   \\ 
& here     & $0.12\pm 0.01$ & $1.84\pm 0.03$ & $2.74\pm 0.01$ & $0.97$   \\ 
\hline
\multirow{2}{25mm}{$3\Mpc$ top-hat} 
& NFW      & $0.10\pm 0.01$ & $1$            & $3$            & $0.98$   \\ 
& here     & $0.07\pm 0.01$ & $1.5 \pm 0.2 $ & $2.46\pm 0.01$ & $0.98$   \\ 
\hline
\multirow{2}{25mm}{$(7\Mpc)^{-1}$ k-cutoff} 
& NFW      & $0.09\pm 0.01$ & $1$            & $3$            & $0.97$   \\ 
& here     & $0.3\pm 0.2$   & $2.1\pm 0.2$   & $2.8\pm 0.2$   & $0.99$   \\ 
\hline
\multirow{2}{25mm}{$7\Mpc$ top-hat} 
& NFW      & $0.28\pm 0.01$ & $1$            & $3$            & $0.99$   \\ 
& here     & $0.23\pm 0.01$ & $1.70\pm 0.03$ & $2.23\pm 0.01$ & $0.98$   \\ 
\hline
\multirow{2}{25mm}{$(14\Mpc)^{-1}$ k-cutoff} 
& NFW      & $0.21\pm 0.01$ & $1$            & $3$            & $0.99$   \\ 
& here     & $0.18\pm 0.01$ & $1.78\pm 0.03$ & $2.28\pm 0.01$ & $0.98$   \\ 
\hline
\multirow{2}{25mm}{low-res. unsmoothed} 
& NFW      & $0.26\pm 0.03$ & $1$            & $3$            & $1.01$   \\ 
& here     & $0.23\pm 0.01$ & $2.08\pm 0.06$ & $2.38\pm 0.05$ & $1.01$   \\ 
\hline
\multirow{2}{25mm}{low-res. $7\Mpc$ top-hat} 
& NFW      & $0.24\pm 0.02$ & $1$            & $3$            & $1.00$   \\ 
& here     & $0.26\pm 0.1 $ & $4\pm 2$       & $2.4\pm 0.1$   & $1.40$   \\ 
\hline
\end{tabular}
\end{center}
\caption[Fits to the dark matter density profiles of all the models]{
Coefficients for fits to the dark matter density profiles for all the models.
}
\label{tab.Res.Prof.Form.coef.all}
\end{table}

\begin{table}
\begin{center}
\begin{tabular}{lcc}
\hline
			& $\tau_{num}$   & $n$            \\
\hline
unsmoothed		& $5.57\pm 0.50$ & $0.61\pm 0.03$ \\
$3\Mpc$ top-hat		& $5.89\pm 0.61$ & $0.67\pm 0.04$ \\
$(7\Mpc)^{-1}$ freq. cut& $6.96\pm 0.63$ & $0.75\pm 0.04$ \\
$7\Mpc$ top-hat		& $4.80\pm 0.86$ & $0.67\pm 0.12$ \\
$(14\Mpc)^{-1}$ freq. cut& $6.96\pm 0.63$ & $0.75\pm 0.04$ \\
low-res. unsmoothed	& $5.7 \pm 0.5 $ & $0.66\pm 0.07$ \\
low-res $7\Mpc$ top-hat	& $5.7 \pm 1.5 $ & $0.68\pm 0.26$ \\
\hline
\end{tabular}
\end{center}
\caption[Mass-temperature scaling law with free coefficients]{
Mass-temperature scaling law with free coefficients.}
\label{tab.Res.MT.freecoef}
\end{table}

\begin{table}
\begin{center}
\begin{tabular}{lccc}
\hline
			& $\tau_{num}$ \\
\hline
unsmoothed		& $6.8\pm 1.6$ \\
$3\Mpc$ top-hat		& $6.2\pm 1.1$ \\
$(7\Mpc)^{-1}$ k-cutoff	& $6.2\pm 1.1$ \\
$7\Mpc$ top-hat		& $4.9\pm 0.9$ \\
$(14\Mpc)^{-1}$ k-cutoff& $5.5\pm 1.1$ \\
low-res. unsmoothed	& $5.8\pm 1.3$ \\
low-res. $7\Mpc$ top-hat& $4.9\pm 1.0$ \\
\hline
\end{tabular}
\end{center}
\caption[Mass-temperature scaling law coefficient]{
Mass-temperature scaling law coefficient. The coefficient, $\tau_{num}$, of \eq{eq.Res.MT.Tr} is given for the simulations.  It relates the cluster temperature (in $\keV$) with the cluster mass (in $10^{15}\Msun$).}
\label{tab.Res.MT.coef}
\end{table}

\begin{table}
\begin{center}
\begin{tabular}{ccccc}
                         & $r_s$          & $\alpha$       & $\beta$        & $\chi^2/N$ \\ 
\hline
unsmoothed               & $0.43\pm 0.03$ & $2.37\pm 0.01$ & $2.45\pm 0.01$ & $0.96$     \\ 
$3\Mpc$ top-hat          & $0.56\pm 0.03$ & $2.27\pm 0.01$ & $2.80\pm 0.04$ & $0.93$     \\ 
$(7\Mpc)^{-1}$ k-cutoff  & $0.54\pm 0.01$ & $2.24\pm 0.01$ & $2.97\pm 0.02$ & $0.94$     \\ 
$7\Mpc$ top-hat          & $0.59\pm 0.01$ & $1.60\pm 0.02$ & $2.99\pm 0.03$ & $1.00$     \\ 
$(14\Mpc)^{-1}$ k-cutoff & $0.23\pm 0.01$ & $1.78\pm 0.03$ & $2.34\pm 0.01$ & $0.94$     \\ 
low-res. unsmoothed      & $0.43\pm 0.05$ & $3   \pm 2   $ & $2.34\pm 0.03$ & $1.00$     \\ 
low-res. $7\Mpc$ top-hat & $0.73\pm 0.03$ & $1.54\pm 0.08$ & $3.0 \pm 0.1 $ & $1.01$     \\ 
\hline
\end{tabular}
\end{center}
\caption[Coefficients for fits to the gas density profiles]{
Coefficients for fits to the gas density profiles. All use the discontinuous form.
}
\label{tab.Res.Prof.Gas.Rho}
\end{table}

\begin{table}
\begin{center}
\begin{tabular}{ccccc}
                         & $r_s$            & $\beta$          & $\chi^2/N$ \\
\hline
unsmoothed               & $0.020\pm 0.001$ & $0.808\pm 0.001$ & $0.96$     \\
$3\Mpc$ top-hat          & $0.044\pm 0.001$ & $0.840\pm 0.002$ & $0.94$     \\
$(7\Mpc)^{-1}$ k-cutoff  & $0.056\pm 0.001$ & $0.868\pm 0.002$ & $0.95$     \\
$7\Mpc$ top-hat          & $0.145\pm 0.005$ & $0.811\pm 0.001$ & $1.04$     \\
$(14\Mpc)^{-1}$ k-cutoff & $0.073\pm 0.002$ & $0.791\pm 0.004$ & $0.94$     \\
low-res. unsmoothed      & $0.02 \pm 0.02 $ & $0.78 \pm 0.02 $ & $4.05$     \\
low-res. $7\Mpc$ top-hat & $0.42 \pm 0.04 $ & $1.05 \pm 0.05 $ & $1.02$     \\
\hline
\end{tabular}
\end{center}
\caption[Coefficients for the $\beta$-fit to the gas density profiles]{
Coefficients for the $\beta$-fit to the gas density profiles.
}
\label{tab.Res.Prof.Gas.Rho.Beta}
\end{table}

\begin{table}
\begin{center}
\begin{tabular}{ccccc}
                         & $\tau_o$       & $R_{iso}$      & $-n_T$         & $\chi^2/N$ \\
\hline
unsmoothed               & $8.5 \pm 1   $ & $0.10\pm 0.03$ & $0.41\pm 0.04$ & $1.02$     \\
$3\Mpc$ top-hat          & $8.6 \pm 0.1 $ & $0.10\pm 0.03$ & $0.47\pm 0.01$ & $1.01$     \\
$(7\Mpc)^{-1}$ k-cutoff  & $7.52\pm 0.04$ & $0.16\pm 0.01$ & $0.57\pm 0.01$ & $1.03$     \\
$7\Mpc$ top-hat          & $4.85\pm 0.04$ & $0.26\pm 0.01$ & $0.56\pm 0.02$ & $1.01$     \\
$(14\Mpc)^{-1}$ k-cutoff & $6.59\pm 0.04$ & $0.18\pm 0.01$ & $0.57\pm 0.01$ & $1.06$     \\
low-res. unsmoothed      & $9.9 \pm 0.5 $ & $0.5 \pm 0.1 $ & $0.7 \pm 0.2 $ & $1.03$     \\
low-res. $7\Mpc$ top-hat & $4.0 \pm 0.1 $ & $0.58\pm 0.02$ & $0.92\pm 0.07$ & $1.00$     \\
\hline
\end{tabular}
\end{center}
\caption[Coefficients for fits to the gas temperature profiles]{
Coefficients for fits to the scaled gas temperature profiles. All use a discontinuous form with an iso-thermal core.}
\label{tab.Res.Prof.Gas.T}
\end{table}

\begin{table}
\begin{center}
\begin{tabular}{lccc}
\hline
                        & $\tau_{analy}$ & $\tau_{num}$ & $\tau_o$       \\
\hline
unsmoothed 		& $7.7\pm 0.2$   & $6.8\pm 1.6$ & $8.5 \pm 1   $ \\
$3\Mpc$ top-hat		& $6.8\pm 0.2$   & $6.2\pm 1.1$ & $8.6 \pm 0.1 $ \\
$(7\Mpc)^{-1}$ k-cutoff	& $7.1\pm 0.3$   & $6.2\pm 1.1$ & $7.52\pm 0.04$ \\
$7\Mpc$ top-hat		& $6.4\pm 0.7$   & $4.9\pm 0.9$ & $4.85\pm 0.04$ \\
$(14\Mpc)^{-1}$ k-cutoff& $7.1\pm 0.3$   & $5.5\pm 1.1$ & $6.59\pm 0.04$ \\
low-res. unsmoothed	& $7.6\pm 0.4$   & $5.8\pm 1.3$ & $9.9 \pm 0.5 $ \\
low-res. $7\Mpc$ top-hat& $5.4\pm 0.4$   & $4.9\pm 1.0$ & $4.0 \pm 0.1 $ \\
\hline
\end{tabular}
\end{center}
\caption[Analytic and numeric coefficients for the M-T relation]{
Analytic and numeric coefficients for the M-T relation.  Given are the semi-analytic estimates for the scaled temperature parameter, $\tau_{analy}$, the numerical value from fitting to the temperature-mass distribution, $\tau_{num}$, and the values derived from fitting the mean profiles, $\tau_o$.}
\label{tab.Res.Tau.coef}
\end{table}

\renewcommand{\thisdir}{Figures}

\renewcommand{\figscale}{0.7}
\begin{figure}
\epsscale{\figscale}
\plotone{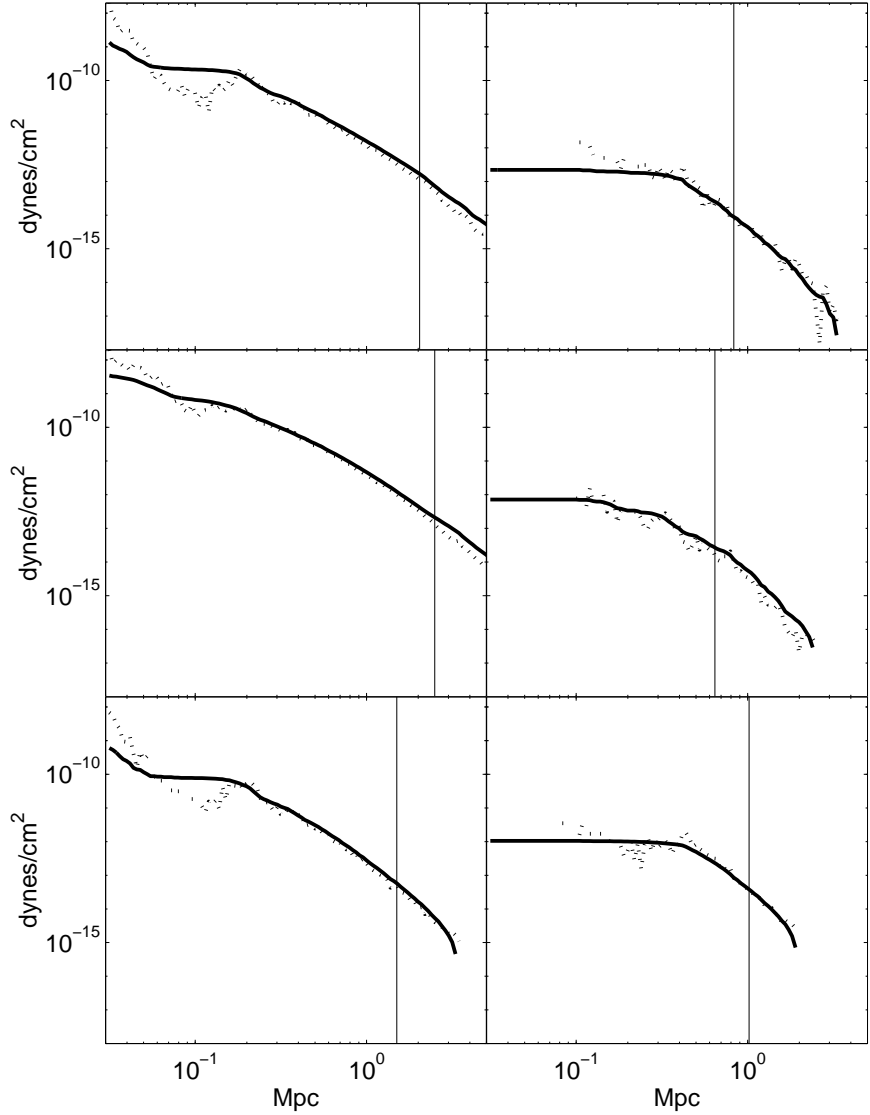}
\caption{
Comparison of the gas pressure profile in the simulated clusters to that  expected for a hydrostatic scenario for those clusters formed hierarchically.  Plotted are the pressure profiles for the clusters with the three highest central pressures (left) and the three lowest central pressures (right).  The solid line is calculated from the integration of the equation of hydrostatic equilibrium while the dotted line is the measured gas pressure profile.  The vertical line denotes the virial radius, $R_{200}$.
}
\label{fig.hydrostatic.hydro_eq.run101}
\end{figure}

\begin{figure}
\epsscale{\figscale}
\plotone{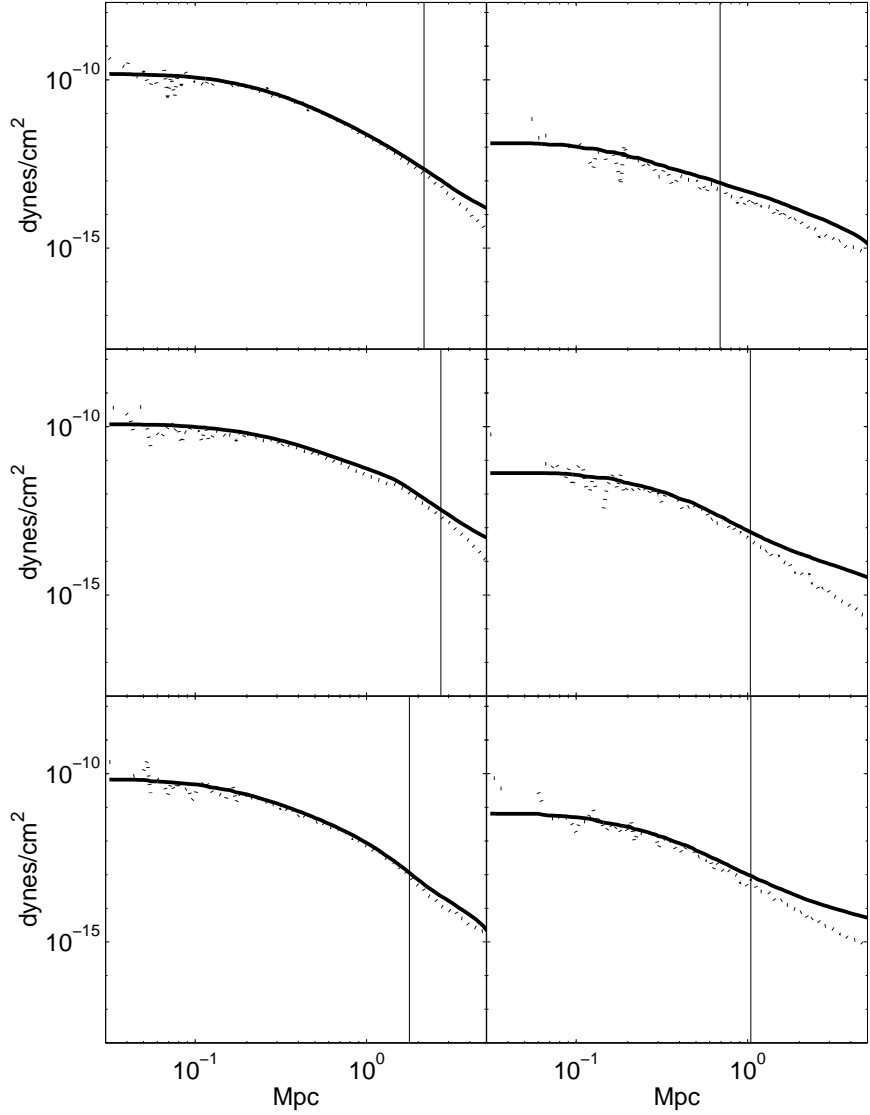}
\caption{
Comparison of the measured gas pressure profile to that expected for a hydrostatic cluster formed non-hierarchically.  Plotted are the pressure profiles for six of the seven clusters in the sample.  As in \fig{fig.hydrostatic.hydro_eq.run101}, the data for the three clusters with the highest central pressures are on the left, and those for the three lowest central pressures are on the right.  The solid line is calculated from the integration of the equation of hydrostatic equilibrium while the dotted line is the actual gas pressure profile.  The vertical line denotes $R_{200}$, the approximate virial radius.
}
\label{fig.hydrostatic.hydro_eq.run103}
\end{figure}

\renewcommand{\figscale}{1}
\begin{figure}
\epsscale{\figscale}
\plotone{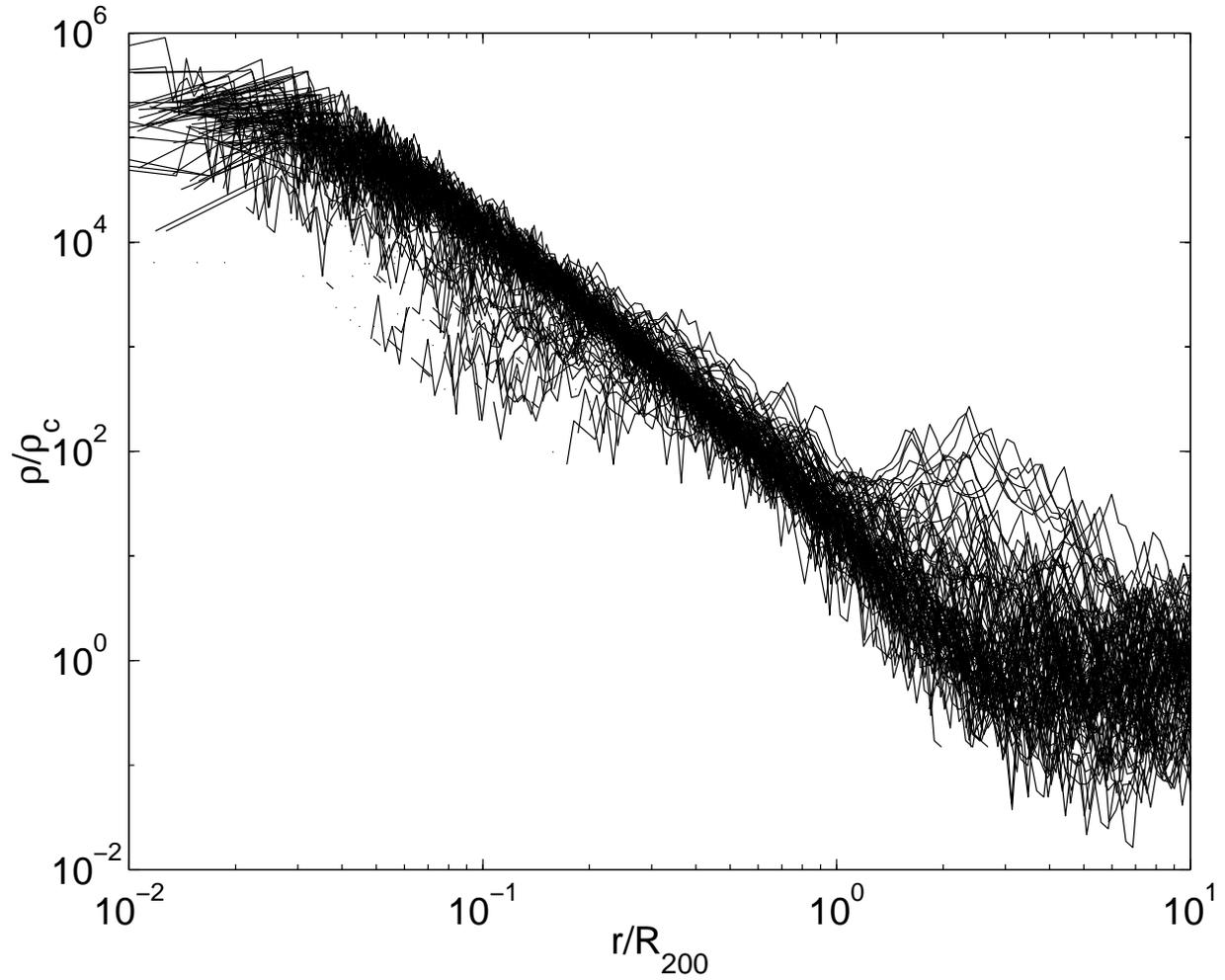}
\caption{
Dark matter density profiles for the clusters formed hierarchically at the epoch corresponding to the present.  The radii have been scaled to $R_{200}$.}
\label{fig.Res.Prof.allDM}
\end{figure}

\begin{figure}
\epsscale{\figscale}
\plotone{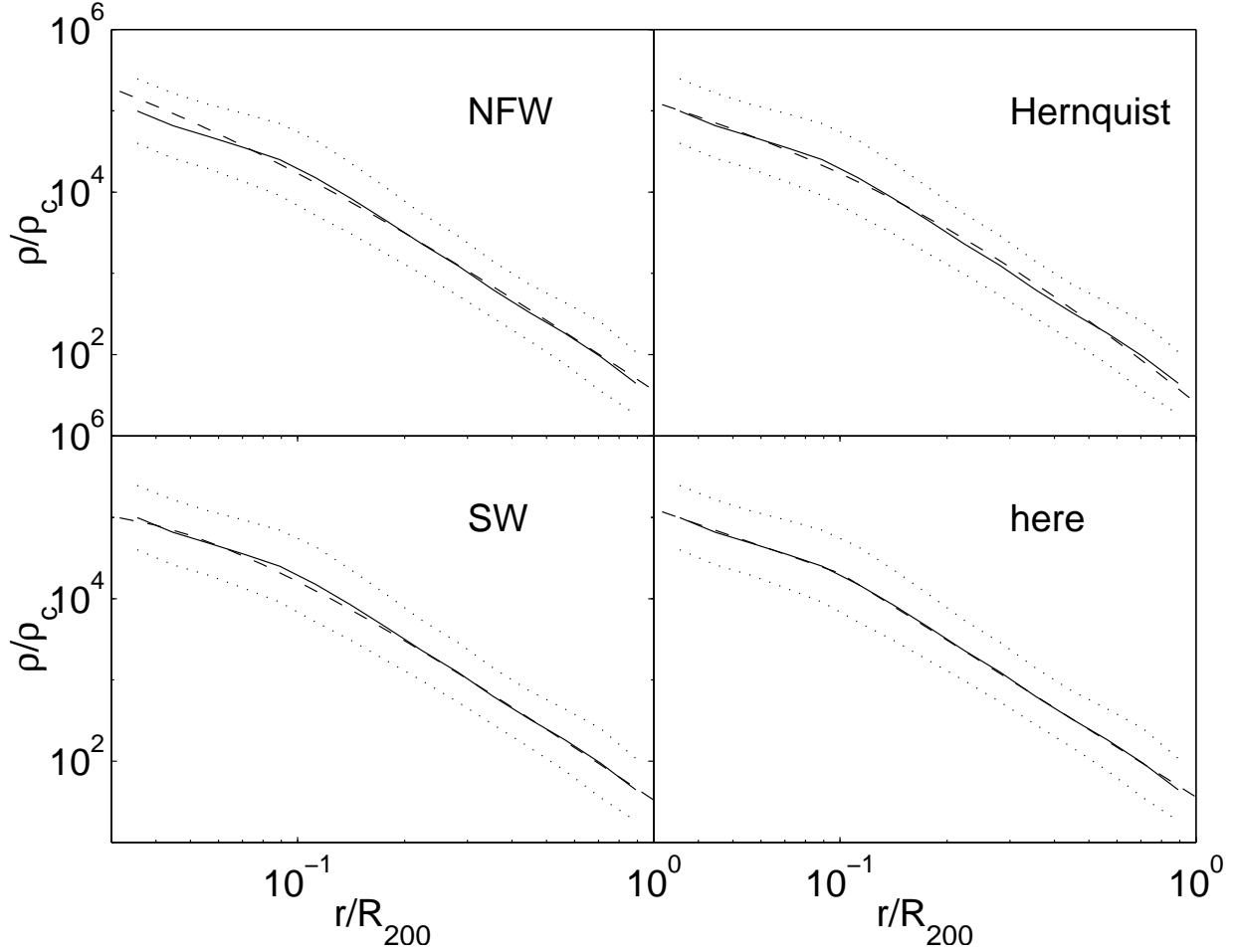}
\caption{
Hierarchical data. Fits to the mean dark matter density profile for the forms of Navarro, Frenk, and White (top left), Hernquist (top right), Syer and White (lower left), and the form suggested here (lower right). The dashed lines are the fits. The solid lines are the mean profiles with $1\sigma$ variations bound by the dotted lines.
}
\label{fig.Res.Prof.Form.fits.hier}
\end{figure}

\begin{figure}
\epsscale{\figscale}
\plotone{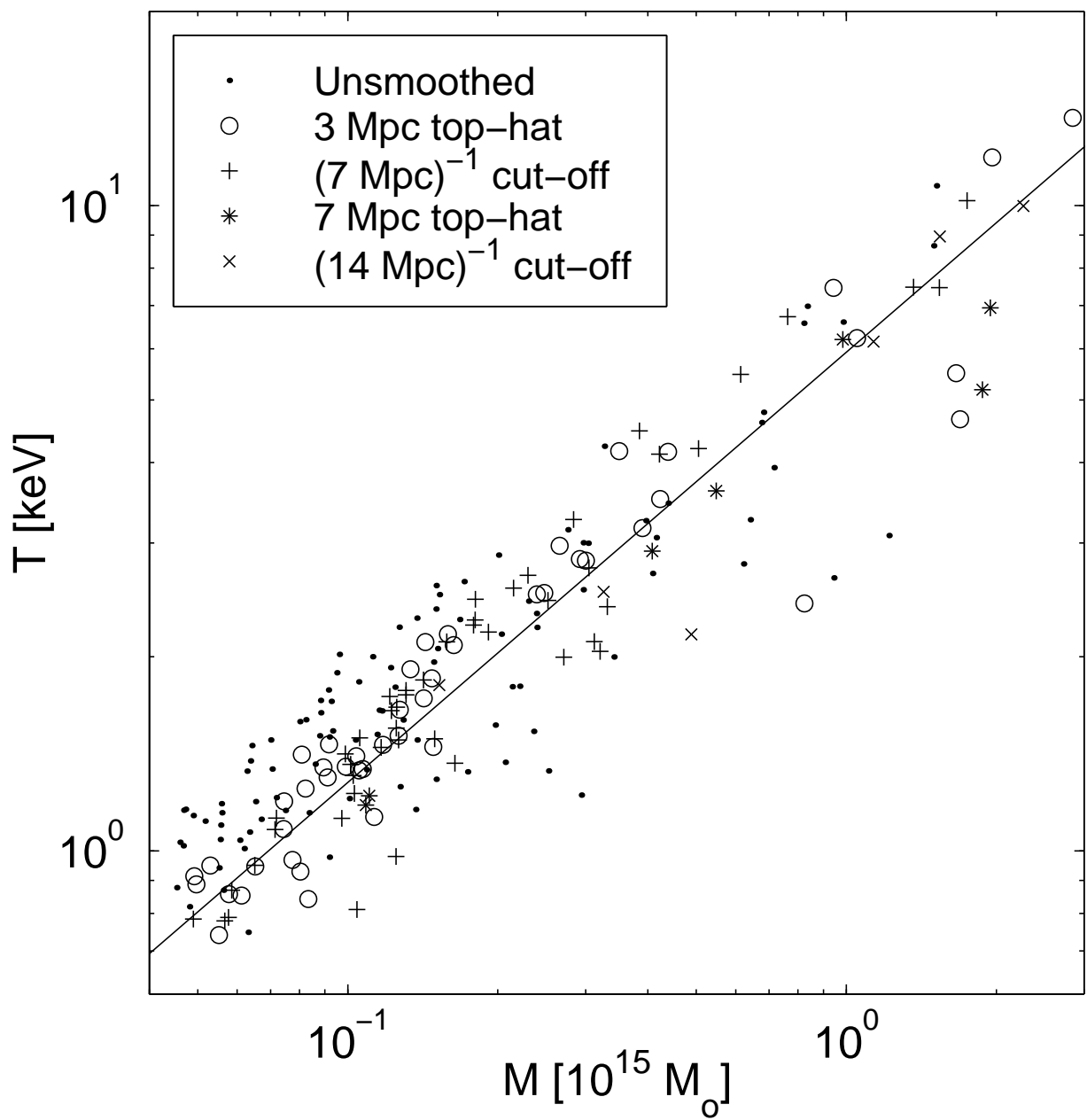}
\caption[Temperature {\it vs} mass]{
The temperature {\it versus} mass for the clusters. The solid line is the fit $T=5.9M^{2/3}$.}
\label{fig.MT.TvsM}
\end{figure}

\begin{figure}
\epsscale{\figscale}
\plotone{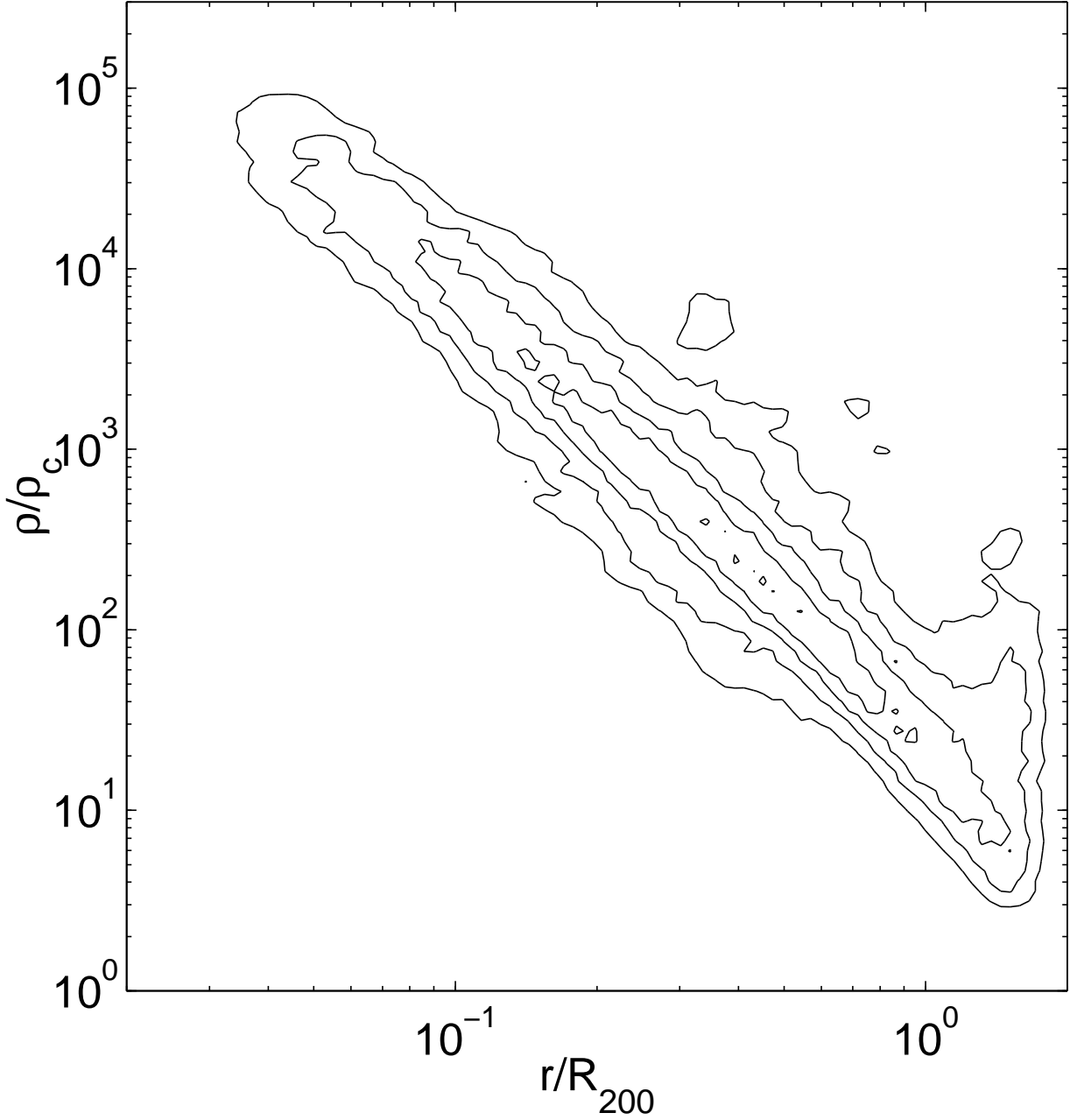}
\caption{Gas densities for the hierarchical run.  The densities are for the sample of particles described in the text. The radii have been scaled to $R_{200}$.  The total sample corresponds to about 30\% of all gas particles in the simulation. Contour lines are of constant mass surface density in $r/R_{200}-T$ space, separated by factors of $10^{\onehalf}$.}
\label{fig.MT.GasDensity.101}
\end{figure}

\begin{figure}
\epsscale{\figscale}
\plotone{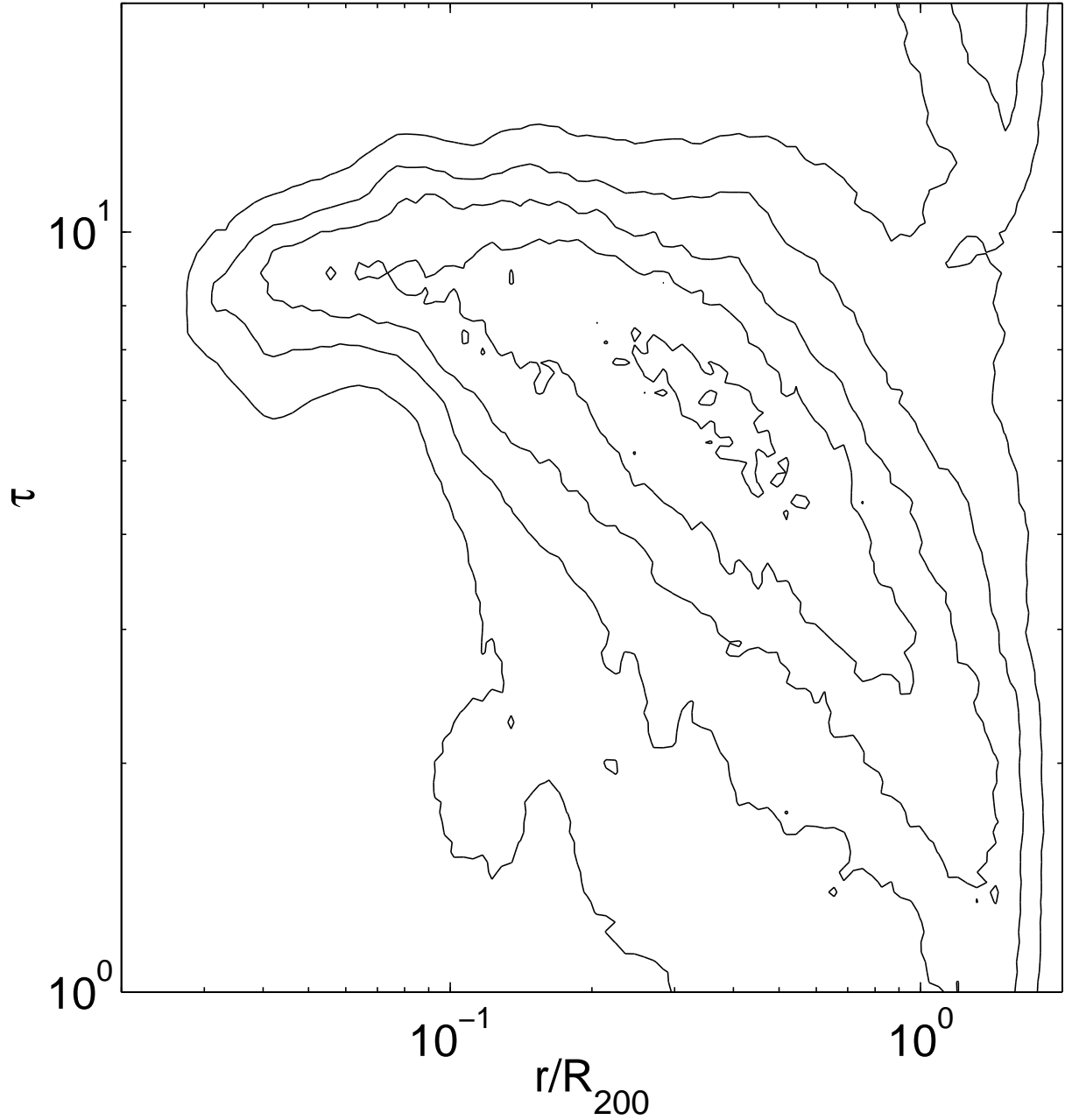}
\caption{The scaled-temperature parameter, $\tau$, for the hierarchical run.  The sample of particles is the same as in \fig{fig.MT.GasDensity.101}. Contour lines are of constant mass surface density in $r/R_{200}-\tau$ space, separated by factors of $10^{\onehalf}$.}
\label{fig.MT.GasTau.101}
\end{figure}

\end{document}